\documentclass[aps,pre,reprint,amsmath,amssymb,onecolumn,nofootinbib]{revtex4-2}

\usepackage{graphicx}
\usepackage{bm}
\renewcommand{\vec}[1]{\boldsymbol{#1}}

\def\cpc{{\itshape Comput. Phys. Commun.}, }
\def\jgr{{\itshape J. Geophys. Res.}, }
\def\apj{{\itshape Astrophys. J.}, }
\def\apjs{{\itshape Astrophys. J. Suppl.}, }
\def\grl{{\itshape Geophys. Res. Lett.}, }
\def\mnras{{\itshape Mon. Notices Royal Astron. Soc.}, }
\def\pop{{\itshape Phys. Plasmas}, }
\def\prl{{\itshape Phys. Rev. Lett.}, }
\def\ssr{{\itshape Space Sci. Rev.}, }

\begin{document}

\title{Loading non-Maxwellian Velocity Distributions in Particle Simulations}

\author{Seiji Zenitani}
\affiliation{Space Research Institute, Austrian Academy of Sciences, 8042 Graz, Austria}
\email{seiji.zenitani@oeaw.ac.at}

\author{Shunsuke Usami}
\affiliation{National Institute for Fusion Science, Toki 509-5292, Japan}

\author{Shuichi Matsukiyo}
\affiliation{Faculty of Engineering Sciences, Kyushu University 6-1 Kasuga-Koen, Kasuga, Fukuoka 816-8580, Japan}

\begin{abstract}
Numerical procedures for generating
non-Maxwellian velocity distributions in particle simulations are presented.
First, Monte Carlo methods for the $(r,q)$ distribution
that generalizes flattop and Kappa distributions are discussed.
Then, two rejection methods for the regularized Kappa distribution are presented,
followed by a comparison in the $\kappa$ space.
A simple recipe is proposed for the subtracted Kappa distribution.
Properties and numerical recipes for
the ring and shell distributions with a finite Gaussian width are discussed.
The ring and shell Maxwellians are further introduced
as alternatives to the ring and shell distributions.
Finally, methods for the super-Gaussian and the filled-shell distributions are presented.
\end{abstract}

\maketitle

\section{Introduction}

Plasma velocity distribution functions (VDFs) show many different shapes in the heliosphere. 
Since the mean free path is much longer than the typical length scale,
local VDFs deviate significantly from a Maxwell distribution. 
In the presence of a planetary dipole field, the VDF is often loss-cone-shaped,
because particles with field-aligned velocities escape to the planet. 
The loss-cone VDF leads to pressure anisotropy,
which in turn generates various plasma waves in planetary magnetospheres.
VDFs sometimes exhibit a power-law tail in the high-energy part.
Such VDFs are often approximated by a generalized Lorentzian distribution,
now widely known as the Kappa distribution \citep{olbert68,vas68}.
The Kappa distribution has drawn significant attention in plasma physics,
because it might be an outcome of long-range interaction between charged particles \citep{kappa,lazar21,tsallis23}. 
Extensions of the Kappa distribution have been proposed and actively studied,
such as the Kappa distribution with a high-energy cutoff and the distribution with a loss-cone \citep{scherer17,summers91,summers25}.
In the solar wind, electron VDFs often have
Kappa-type and other subcomponents \citep{wilson19}.
In addition, ion VDFs often contain
a ring- or shell-shaped ``pickup'' component
that originates from interstellar neutral atoms \citep{moebius85,zirnstein22}.
A shell-shaped VDF further evolves to a filled-shell VDF,
due to an adiabatic expansion of the solar wind.
In the presence of pickup ions (PUIs), 
local plasma processes can be substantially modified \citep{matsukiyo14,nakanotani23,pyakurel25}.
Associated with electron heating,
flattop-type VDFs are often observed in and around the magnetosphere \citep{feldman82,thomsen83,qureshi14,oka22,richard25}.
Indeed, a wide variety of VDFs are recognized throughout the heliosphere.

Particle-in-cell (PIC) simulation \citep{birdsall,hockney} is
very useful for exploring kinetic processes in plasma systems. 
Without PIC simulation, it is virtually impossible to accurately predict
wave-particle interaction, pitch-angle scattering, and particle acceleration
in the aforementioned VDFs. 
For a Maxwell distribution,
one can use a built-in procedure to generate a normal distribution or 
one can use the \citet{bm58} method. 
However, it is not always clear
how to generate non-Maxwellian VDFs in simulation. 
For example, despite its importance, numerical procedures
for Kappa distributions have been largely unknown to the community,
until \citet{abdul14,abdul15} used its relation to Student's $t$ distributions. 
More recently, \citet{zeni22,zeni23} proposed numerical procedures for
Kappa, relativistic, and loss-cone distributions. 
Their methods rely on only three types of basic random numbers:
uniform, normal, and gamma random numbers.

Yet, there are many other VDFs that simulation scientists would like to use,
as discussed later in this article. 
To generate such distributions,
one may consider general-purpose methods such as
acceptance-rejection sampling or inverse transform sampling. 
In the acceptance-rejection method,
we need to define a good envelope function
to make the rejection process efficient;
otherwise the algorithm becomes increasingly inefficient,
especially in three dimensions. 
The inverse transform method relies on
the inverse of the cumulative distribution function (CDF),
which is not always available analytically.
Moreover, it is difficult to extend the method to multiple dimensions. 
Considering the limitations of these general-purpose methods,
it is necessary to develop dedicated numerical procedures
for a variety of non-Maxwellian VDFs in heliophysics.

The purpose of this paper is to comprehensively present
numerical procedures for generating non-Maxwellian VDFs in particle simulations,
that are not covered by our recent studies \citep{zeni22,zeni23}. 
We provide ``numerical recipes'' for nine different distributions,
in the hope that
they make kinetic modeling of plasma processes easier than before. 
The rest of the paper is organized as follows.
In Section \ref{sec:prep}, we summarize random number generators
for basic probability distributions in statistics. 
Section \ref{sec:rq} discusses
the $(r,q)$ distribution that generalizes the Kappa and flattop distributions.
A gamma-distribution method and a piecewise rejection method are evaluated in the 2-D parameter space.
Section \ref{sec:RKD} discusses the regularized Kappa distribution.
Two rejection methods, a post-rejection method and a piecewise rejection method are proposed. 
Section \ref{sec:SK} develops a procedure for a subtracted Kappa distribution,
a recently-proposed Kappa loss-cone (KLC) model with a narrow loss cone.
Section \ref{sec:ring} discusses ring and shell distributions with finite Gaussian width.
Two methods are briefly presented.
Section \ref{sec:ringMax} presents ring and shell Maxwellians.
In particular, we introduce a shell Maxwell distribution
as a handy alternative to the conventional shell distribution.
Section \ref{sec:other} presents other isotropic distributions.
The super Gaussian distribution and the filled-shell distributions are discussed.
Section \ref{sec:verification} validates numerical accuracies of the proposed methods for all the distributions. 
Section \ref{sec:discussion} contains a discussion and summary. 

\section{Basic statistical distributions}
\label{sec:prep}

\subsection{Exponential distribution}
\label{sec:exp}

The exponential distribution is defined in $0 \le x<\infty$.
Its probability density is given by
\begin{align}
\mathrm{Exp}(x; \lambda) &=
\frac{ 1 }{ \lambda }
\exp
\left(
-\frac{x}{\lambda} \right)
\end{align}
Here the parameter $\lambda$ controls the scale length in $x$.
As well known, a random number that follows the exponential distribution
can be obtained by $- \lambda \log U$,
by using a uniform random number $U \in (0,1]$.
One can use $- \lambda \log (1-U)$ instead,
when $U$ is defined in $[0,1)$.

\subsection{Gamma distribution}
\label{sec:gamma}

The gamma distribution is one of the most important probability distributions.
The probability density $\mathrm{Ga}(x)$ is defined in $0 \le x<\infty$:
\begin{align}
\mathrm{Ga}(x; k, \lambda) &=
\frac{ x^{k-1}  }{ \Gamma(k) \lambda^{k}}
\exp
\left(
-\frac{x}{\lambda} \right)
\end{align}
Here, $\Gamma(x)$ is the gamma function.
The distribution is parameterized by the shape parameter $k$ and
the scale parameter $\lambda$. 
The exponential distribution is a special case of the gamma distribution,
i.e., $\mathrm{Exp}(x; \lambda) = \mathrm{Ga}(x; 1, \lambda)$.
To generate a random number drawn from a gamma distribution,
which we call a gamma variate,
we may use random number generators offered by the system. 
If gamma generators are unavailable or if we want to make our code portable,
we can implement gamma generators by using state-of-art algorithms.
Depending on the shape parameter $k$, we have the following choices:
\begin{itemize}
\item
For $k>1$, \citet{mt00}'s algorithm is the de-facto standard. There are simpler methods for integer and half-integer cases (\citet{kroese11,yotsuji10}. See also \citet[][Appendix A]{zeni22}).
\item
For $k=1$, the distribution is equivalent to the exponential distribution.
We can use the simple procedure, as discussed in Sec.~\ref{sec:exp}.
\item
For $k<1$, several acceptance-rejection methods have been developed \citep[e.g.,][]{AD74,best83}. Based on our experience, we recommend either \citet[][Chapter VII, Sec. 2.6]{devroye86} or \citet{zeni24b} methods.
\end{itemize}
In this paper, we denote $X_{\mathrm{Ga}(k,\lambda)}$ as a gamma variate with the parameters $k, \lambda$.

\subsection{Generalized beta-prime distribution}
\label{sec:betap}

The generalized beta-prime distribution is defined in the following way.
The four parameters need to be positive.
\begin{align}
\label{eq:gbetap}
\mathrm{B'}(x;\alpha,\beta,\gamma,\delta)
&=
\frac{\gamma \Gamma(\alpha+\beta)}{\delta^{\alpha \gamma} \Gamma(\alpha)\Gamma(\beta)}
\left( 1 + \left(\frac{x}{\delta}\right)^\gamma \right)^{-(\alpha+\beta)}x^{\alpha \gamma - 1}
\end{align}
A beta-prime variate can be obtained from two gamma variates \citep[][Section III]{zeni22}.
\begin{align}
\label{eq:gbetap_rand}
X_{\mathrm{B'}(\alpha,\beta,\gamma,\delta)}
=
\delta
\left(
\frac{
X_{\mathrm{Ga}(\alpha,\epsilon)}
}{
X_{\mathrm{Ga}(\beta,\epsilon)}
}
\right)^{1/\gamma}
\end{align}

\subsection{Student's $t$ distribution}
\label{sec:t}

Student's $t$ distribution is defined by
\begin{align}
\mathrm{St}(\vec{x}; \nu, \sigma, p)d\vec{x}
=& \frac{1}{(\pi \nu\sigma^2)^{p/2}}
\frac{\Gamma[(\nu+p)/2]}{\Gamma(\nu/2)}
\left( 1 + \frac{ 1 }{\nu} \frac{\vec{x}^2}{\sigma^2} \right)^{-\frac{(\nu+p)}{2}} d\vec{x}
\label{eq:multi_t}
\end{align}
where $\nu$ is degree of freedom, $p$ is the dimension, and
$\vec{x}$ is a multi-dimensional ($p$-dimensional) variable.
Such a $t$-variate $\vec{x}$ can be obtained by
\begin{align}
\vec{x} = \sqrt{\nu\sigma^2} \frac{\vec{n}}{\sqrt{X_{\mathrm{Ga}(\nu/2,2)}}},
\label{eq:t_gen}
\end{align}
where $\vec{n}$ is a $p$-dimensional normal variate.
Eq.~\eqref{eq:t_gen} is identical to a well-known procedure with a chi-squared variate \citep{yotsuji10,kroese11},
because the gamma distribution $\mathrm{Ga}(\nu/2,2)$ is equivalent to the chi-squared distribution with $\nu$ degrees of freedom.

\section{$(r,q)$ distribution}
\label{sec:rq}

\begin{figure*}[htbp]
\centering
\includegraphics[width={\textwidth}]{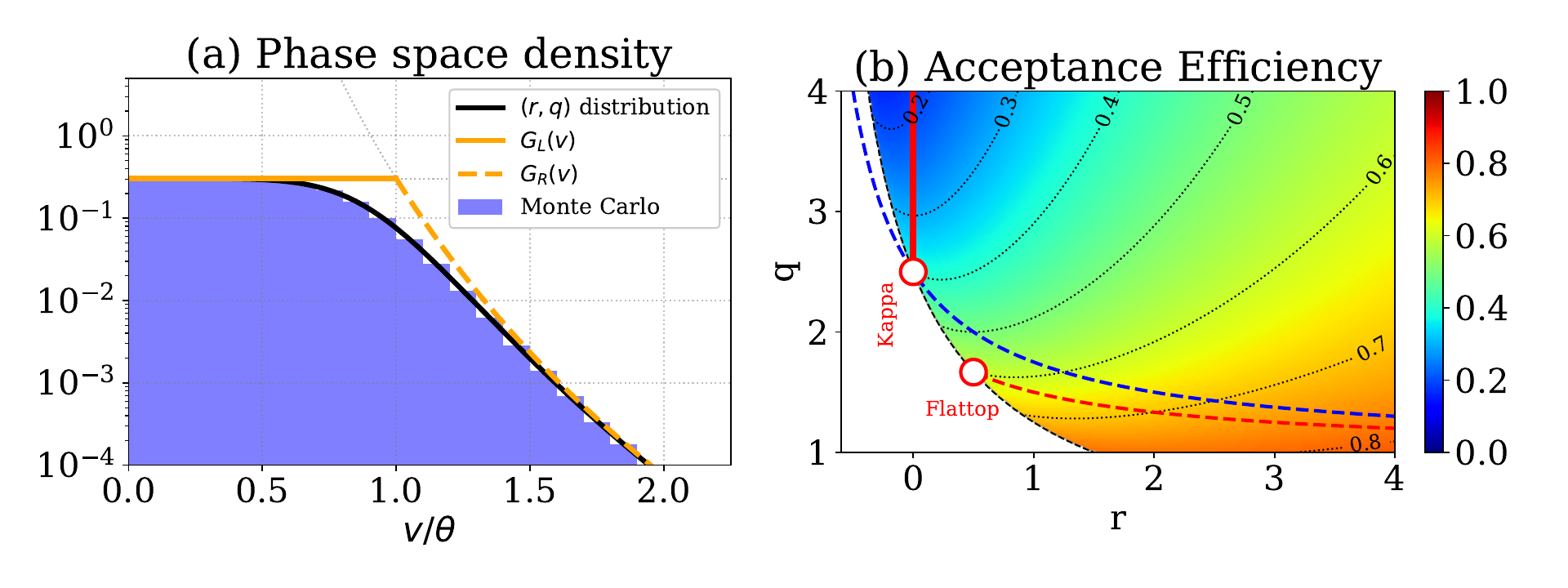}
\caption{
(a)
Phase space density of the $(r,q)$ distribution with $(r,q,\theta_\parallel, \theta_\perp)=(2,2,\theta,\theta)$.
The distribution function (black line; Eq.~\eqref{eq:rq}) and normalized Monte Carlo results (blue histogram) of $10^6$ particles.
The solid and dashed orange lines show the left and right envelope functions (Eq. \eqref{eq:rq_env}).
(b)
Acceptance efficiency for the $(r,q)$ distribution (Eq.~\eqref{eq:rq_eff}).
The vertical red line corresponds to the Kappa distribution and
the red dashed line corresponds to the flattop distribution.
Below the blue dashed line,
the beta-prime method requires 
a gamma distribution with a shape parameter less than unity in the denominator.
\label{fig:rq}}
\end{figure*}

The generalized $(r,q)$ distribution was proposed to model
flattop-type non-Maxwellian VDFs \citep{zaheer04,qureshi04}.
It often draws attention
in and around the terrestrial magnetosphere \citep[e.g.,][]{qureshi14,qureshi19,richard25}.
The phase space density of the distribution is defined as follows:
\begin{align}
f_\mathrm{rq}(\vec{v}; r, q, \theta_\parallel, \theta_\perp) d^3v
&=
N_0\cdot
C_\mathrm{rq}
\left( 1 + \frac{1}{q-1} \left( \frac{ v_\parallel^2 }{ \theta_\parallel^2 } + \frac{ v_\perp^2 }{ \theta_\perp^2 } \right)^{1+r} \right)^{-q}
 d^3v,
\label{eq:rq}
\\
C_\mathrm{rq}
&=
\frac{3 \Gamma(q)}{4\pi (q-1)^{\frac{3}{2(1+r)}} \theta_\parallel \theta_\perp^2
\Gamma\left(1+\frac{3}{2(1+r)}\right)\Gamma\left(q-\frac{3}{2(1+r)}\right)}
,
\label{eq:Crq}
\end{align}
where $N_0$ is the plasma number density,
$C_\mathrm{rq}$ is the normalization constant, and
$\theta_\parallel$ and $\theta_\perp$ are characteristic speeds
in the parallel and perpendicular directions with respect to the magnetic field.
We consider $r > -1$ and $q>1$. 
The pressures and the kinetic energy density of the distribution are
\begin{align}
P_\parallel
&=
\frac{1}{3} N_0 m \theta_\parallel^2
K_\mathrm{rq},
~~~
P_\perp
=
\frac{1}{3} N_0 m \theta_\perp^2
K_\mathrm{rq},
~~~~
\mathcal{E} = 
\frac{1}{6} N_0 m \left( \theta_\parallel^2 + 2\theta_\perp^2 \right) K_\mathrm{rq},
\\
K_\mathrm{rq} &= 
(q-1)^{\frac{1}{(1+r)}}
\frac{\Gamma\left(\frac{5}{2(1+r)}\right)\Gamma\left(q-\frac{5}{2(1+r)}\right)}
{\Gamma\left(\frac{3}{2(1+r)}\right)\Gamma\left(q-\frac{3}{2(1+r)}\right)}
\end{align}
To keep these second velocity moments finite,
the parameters must satisfy:
\begin{align}
q - \frac{5}{2(1+r)} > 0.
\label{eq:rq_undefined}
\end{align}
The phase space density (Eq.~\eqref{eq:rq}) with $(r,q,\theta_\parallel, \theta_\perp)=(2,2,\theta,\theta)$ is shown by the black line in Figure \ref{fig:rq}(a). 
The distribution gives a flattop-like picture.

The $(r,q)$ distribution reverts to the (bi-)Kappa distribution \citep{olbert68,vas68,kappa} when $r=0$, $q=\kappa+1$. 
\begin{align}
\label{eq:kappa}
f_{\kappa}(\vec{v}; \kappa, \theta_\parallel, \theta_\perp) d^3v
&=
f_\mathrm{rq}\left(\vec{v}; 0, \kappa+1, \theta_\parallel, \theta_\perp \right) d^3v
\nonumber\\
&=
\frac{1}{(\pi\kappa)^{3/2} {\theta}_\parallel {\theta}_\perp^2} \frac{\Gamma(\kappa+1)}{\Gamma(\kappa-1/2)}  \left( 1 + \frac{ 1 }{\kappa} \left( \frac{ v_\parallel^2 }{ {\theta}_\parallel^2 } + \frac{ v_\perp^2 }{ {\theta}_\perp^2 } \right) \right)^{-(\kappa+1)}
d^3v
\end{align}
The distribution also reverts to
a popular form of the (bi-)flattop distribution
\citep{thomsen83,oka22,zeni24a}
when $r = \kappa-1$ and $q = 1 + 1/\kappa$.
Below, $\bar{\theta} = \kappa^{-1/2\kappa} \theta$ is the characteristic flattop speed.
\begin{align}
\label{eq:FT}
f_\mathrm{ft}(\vec{v}; \kappa, \theta_\parallel, \theta_\perp)d^3{v}
&=
f_\mathrm{rq}\left(\vec{v}; \kappa-1, 1 + {1}/{\kappa}, \theta_\parallel, \theta_\perp \right) d^3v
\nonumber\\
&= \frac{3 N_0}{4\pi \bar{\theta}_\parallel \bar{\theta}_\perp^2} \frac{\Gamma\left(1+\frac{1}{\kappa}\right)}{\Gamma\left(1+\frac{3}{2\kappa}\right)\Gamma\left(1-\frac{1}{2\kappa}\right)}
\left( 1 + \left( \frac{ v_\parallel^2 }{ \bar{\theta}_\parallel^2 } + \frac{ v_\perp^2 }{ \bar{\theta}_\perp^2 } \right)^{\kappa}  \right)^{-({\kappa+1})/{\kappa}} 
d^3{v}
,
\end{align}

Next, we discuss a random number generation for the $(r,q)$ distribution. 
For a moment, we assume $\theta_\parallel=\theta_\perp=\theta$.
To help our discussion, we consider a radial distribution in spherical coordinates,
\begin{align}
F_\mathrm{rq}(v) dv
&=
f_\mathrm{rq}(v)
4\pi v^2 \, dv
\label{eq:rq_rad}
\end{align}
One can translate Eq.~\eqref{eq:rq_rad} into the generalized beta-prime distribution (Eq. \eqref{eq:gbetap}),
\begin{align}
F_\mathrm{rq}(v)
&=
N_0
B'\left(v; \frac{3}{2(1+r)}, q-\frac{3}{2(1+r)},~2(1+r),~(q-1)^{\frac{1}{2(1+r)}}\theta\right)
.
\label{eq:betap}
\end{align}
From Eq. \eqref{eq:gbetap_rand},
we can obtain the radial velocity $v$ by using two gamma variates,
\begin{align}
v=
(q-1)^{\frac{1}{2(1+r)}} \theta
\left(
\frac{
X_{\mathrm{Ga}(\frac{3}{2(1+r)},1)}
}{
X_{\mathrm{Ga}(q-\frac{3}{2(1+r)},1)}
}
\right)^{\frac{1}{2(1+r)}}
=
\theta
\left(
(q-1)
\frac{
X_{\mathrm{Ga}(\frac{3}{2(1+r)},1)}
}{
X_{\mathrm{Ga}(q-\frac{3}{2(1+r)},1)}
}
\right)^{\frac{1}{2(1+r)}}
\label{eq:rq_gammas}
\end{align}
After obtaining $v$, we need to scatter the vector in random directions in 3D,
by using two uniform variates $U_1, U_2 \sim U(0,1)$.
We can easily extend it for $\theta_\parallel \ne \theta_\perp$. 
The procedure is presented in Algorithm 3.1 in Table \ref{table:rq}.
Caution is needed when $q \le 1 + \frac{3}{2(1+r)}$.
In such a case, the gamma distribution in the denominator of Eq. \eqref{eq:rq_gammas}
has the shape parameter $k=q-\frac{3}{2(1+r)}$ less than unity. 
As the gamma distribution with shape less than unity has a non-zero density at $v=0$,
we need to take care of the division by zero in Eq. \eqref{eq:rq_gammas}.

\begin{table}
\centering
\caption{Sampling algorithms for the $(r,q)$ distribution
\label{table:rq}}
\begin{tabular}{l}
\\
\hline
{\bf Algorithm 3.1: Beta-prime method}\\
\hline
generate $X_1 \sim \mathrm{Ga}(\frac{3}{2(1+r)},1)$\\
generate $X_2 \sim \mathrm{Ga}(q-\frac{3}{2(1+r)},1)$\\
generate $U_1, U_2 \sim U(0, 1)$ \\
$x ~~\leftarrow [(q-1)X_1/X_2]^{\frac{1}{2(1+r)}}$ \\
$v_\parallel ~~ \leftarrow ~~  \theta_\parallel x ~ ( 2 U_1 - 1 )$ \\
$v_{\perp 1} \leftarrow 2 \theta_\perp x \sqrt{ U_1 (1-U_1) } \cos(2\pi U_2)$ \\
$v_{\perp 2} \leftarrow 2 \theta_\perp x \sqrt{ U_1 (1-U_1) } \sin(2\pi U_2)$ \\
{\bf return} $v_{\parallel}, v_{\perp 1}, v_{\perp 2}$\\
\hline
{\bf Algorithm 3.2: Piecewise rejection method}\\
\hline
$p_2 \leftarrow \dfrac{3}{2q(1+r)}$,~
$p_1 \leftarrow 1 - p_2$,~
$R \leftarrow (q-1)^{\frac{1}{2(1+r)}}$\\
{\bf repeat} \\
~~~~generate $U_1, U_2 \sim U(0, 1)$ \\
~~~~{\bf if} $U_1 \le p_1$ {\bf then} \\
~~~~~~~~$x \leftarrow (U_1/p_1)^{1/3}$ \\
~~~~~~~~{\bf if} $U_2 < \left( 1 + x^{2(1+r)} \right)^{-q}$, {\bf break} \\
~~~~{\bf else} \\
~~~~~~~~$x \leftarrow ((1-U_1)/p_2)^{\frac{1}{3-2q(1+r)}}$ \\
~~~~~~~~{\bf if} $U_2 < (x^{-2(1+r)} + 1)^{-q}$, {\bf break} \\
~~~~{\bf endif} \\
{\bf end repeat} \\
generate $U_3, U_4 \sim U(0, 1)$ \\
$v_\parallel ~~ \leftarrow ~~ \theta_\parallel R x ~ ( 2 U_3 - 1 )$ \\
$v_{\perp 1} \leftarrow 2 \theta_\perp R x \sqrt{ U_3 (1-U_3) } \cos(2\pi U_4)$ \\
$v_{\perp 2} \leftarrow 2 \theta_\perp R x \sqrt{ U_3 (1-U_3) } \sin(2\pi U_4)$ \\
{\bf return} $v_{\parallel}, v_{\perp 1}, v_{\perp 2}$\\
\hline
\end{tabular}
\end{table}

We present another procedure, the piecewise rejection method,
for generating the $(r,q)$ distribution.
We employ the following piecewise envelope functions.
\begin{align}
G(v)
&=
\left\{
\begin{array}{ll}
G_L(v) = C_\mathrm{rq}
~~
& (0 \le v < v_c)
\\ [8pt]
G_R(v) 
=
C_\mathrm{rq}
\left\{ \dfrac{1}{q-1} \left( \dfrac{v}{\theta} \right)^{2(1+r)} \right\}^{-q}
& (v_c \le v < \infty)
\end{array}
\right.
\label{eq:rq_env}
\end{align}
In Figure \ref{fig:rq}(a),
the solid and dashed orange lines indicate the two envelopes in this case. 
The left and right envelopes meet at $v \equiv v_c = (q-1)^{1/2(1+r)} \theta$.
In the radial direction in spherical coordinates, they give the following two areas,
\begin{align}
S_L &=
\int_0^{v_c} G_L(v) 4\pi v^2\,dv
=
\frac{4\pi C_\mathrm{rq} v_c^3}{3}
\\
S_R &=
\int_{v_c}^{\infty} G_R(v) 4\pi v^2\, dv
=
\frac{4\pi C_\mathrm{rq} v_c^3}{2q(1+r)-3}
.
\end{align}
Using the fact that their ratio is $[2q(1+r)-3]:3$,
we can construct a rejection scheme,
as shown in Algorithm 3.2 (Table \ref{table:rq}). 
We split the logical flow at this ratio.
Then, comparing the distribution function and the envelope functions,
we carry out the rejection procedure.
Trivial modifications for $\theta_\parallel \ne \theta_\perp$ are
also included in the procedure. 
The acceptance efficiency is estimated to be
\begin{align}
\mathrm{eff}_\mathrm{rq}
&= 
\frac{1}{S_L + S_R}
= 
\dfrac{
{
\Gamma\left(1+\frac{3}{2(1+r)}\right)\Gamma\left(1+q-\frac{3}{2(1+r)}\right)}
}{
{\Gamma(1+q)}
}
\label{eq:rq_eff}
\end{align}
The efficiency is shown by a colormap
in the $(r,q)$ parameter space in Figure \ref{fig:rq}(b).
The distribution is undefined in the lower-left empty region,
where Eq.~\eqref{eq:rq_undefined} is not satisfied.
Along the vertical red line,
the $(r,q)$ distribution corresponds to the Kappa distribution (Eq.~\eqref{eq:kappa}).
The dashed red line corresponds to the flattop distribution (Eq.~\eqref{eq:FT}).
The efficiency for the flattop distribution is given by
\begin{align}
\mathrm{eff}_\mathrm{ft}
&= 
\dfrac{
{
\Gamma\left(1+\frac{3}{2 \kappa}\right)\Gamma\left(2-\frac{1}{2 \kappa}\right)}
}{
{\Gamma\left(2+\frac{1}{\kappa}\right)}
}
> \frac{3}{5} ~~~\mathrm{(for~\kappa > 3/2)}
.
\label{eq:ft_eff}
\end{align}

Practically, one can generate the $(r,q)$ distribution
by using either of the two methods. 
The piecewise rejection method can be easily implemented, but
it becomes increasingly inefficient
in a Kappa-like regime in the upper-left in Figure \ref{fig:rq}(b).
In such cases, the beta-prime method is favorable. 
Along the red line, one should use
dedicated methods for the Kappa distribution \citep{abdul15,zeni22,zeni25}. 
Below the blue curve in Figure \ref{fig:rq}(b), $q \le 1 + \frac{3}{2(1+r)}$,
we recommend the piecewise rejection method,
because the beta-prime method may encounter the zero-division problem. 
Using Algorithm 3.1,
we have numerically generated $10^6$ particles that follows
the $(r,q)$ distribution with $(r,q)=(2,2)$.
Their distribution is shown by the blue histogram in Figure \ref{fig:rq}(a).
The numerical results are in excellent agreement with Eq.~\eqref{eq:rq}.

\section{Regularized Kappa distribution}
\label{sec:RKD}

The regularized Kappa distribution \citep{scherer17} is
essentially a Kappa distribution with a high-energy cutoff.
Since higher-order velocity moments no longer diverges,
we can use the kappa index of $0<\kappa<3/2$
that are not accessible in the standard Kappa distribution.
Owing to these and other reasons,
the regularized Kappa distribution has attracted
recent attention \citep{lazar21}.
It is defined in the following way.
\begin{align}
f_\mathrm{rk}(\vec{v};\kappa,\theta,\alpha)
d^3 v
&=
\frac{N_0}{(\pi\kappa\theta^2)^{3/2}{\mathcal{U}\left(\frac{3}{2},\frac{3}{2}-\kappa, \alpha^{2}\kappa\right)}}
\left(1 + \frac{v^2}{\kappa\theta^2}\right)^{-(\kappa+1)}
\exp\left({-\alpha^2 \frac{v^2}{\theta^2}}\right)
d^3 v
\label{eq:RKD}
\end{align}
where $\kappa$ is the spectral index, $\theta$ is the characteristic velocity,
$\alpha$ is a cut-off parameter, and
\begin{align}
\label{eq:KummerU}
\mathcal{U}\left(a,b,z\right) =
\frac{1}{\Gamma\left(a\right)}
\int_{0}^{\infty} x^{a-1} (1+x)^{b-a-1} e^{-zx}  ~dx
\end{align}
is Kummer's U-function (Tricomi function). 
The cut-off parameter is usually set to be small, $\alpha \ll 1$.
For convenience, we limit our attention to $0 \le \alpha < 1$.
The energy density is given by the combination of the U-functions \citep{scherer20}.
\begin{align}
\mathcal{E} =
\int f_\mathrm{rk}(\vec{v};\kappa,\alpha) \left(\frac{1}{2}mv^2\right) d^3v
&=
\frac{3}{4} N_0 m \kappa\theta^2~
\frac{ \mathcal{U}\left(\frac{5}{2},\frac{5}{2}-\kappa, \alpha^{2}\kappa\right)}{{\mathcal{U}\left(\frac{3}{2},\frac{3}{2}-\kappa, \alpha^{2}\kappa\right)}}
\end{align}

\begin{figure*}[htbp]
\centering
\includegraphics[width={\textwidth}]{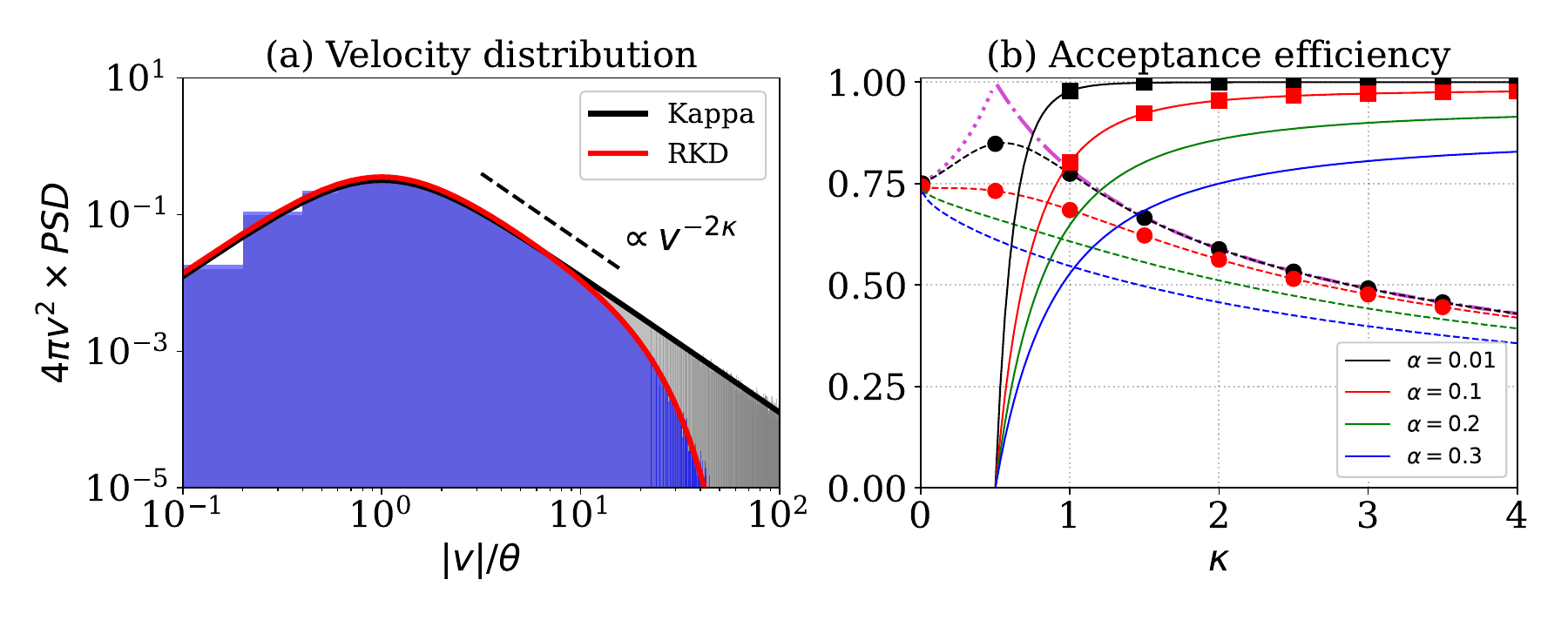}
\caption{
(a) Radial velocity distribution of the regularized Kappa distribution.
The distribution function with $(\kappa, \theta, \alpha)=(1,1,0.05)$ (red line; Eq.~\eqref{eq:RKD}) is compared with numerical results of $10^6$ particles (blue histogram). For reference, the standard Kappa distribution $(\kappa, \theta, \alpha)=(1,1,0)$ (black line; Eq.~\eqref{eq:kappa}) and normalized numerical results of $10^6$ particles (gray histogram) are presented.
(b) Acceptance efficiency. The theory (solid lines; Eq.~\eqref{eq:RKD_eff}) and numerical results (squares) of the post-rejection method, and the theory (dashed lines; Eq.~\eqref{eq:RKD_eff2}) and numerical results (circles) of the piecewise rejection method are presented. The dash-dotted and dotted purple lines indicate upper-bounds (Eqs.~\eqref{eq:RKD_eff3} and \eqref{eq:RKD_eff4}).
\label{fig:RKD}}
\end{figure*}

An omni-directional distribution $4\pi v^2 f_\mathrm{rk}({v})$
for $(\kappa,\theta,\alpha) = (1, 1, 0.05)$
is presented by the red line in Figure \ref{fig:RKD}(a).
The distribution of a standard Kappa distribution with $(\kappa,\theta,\alpha) = (1, 1, 0)$
is also presented by the black line.
One can see power-law slopes of the Kappa distributions, with index $\sim -2\kappa$.
The density of the regularized Kappa distribution drops
beyond the cut-off velocity of $v_c = \theta/\alpha = 20\theta$.
It has three characteristic energy ranges:
the thermal core ($0 < v < \theta$), the power-law slope ($\theta < v \lesssim v_c$),
and the high-energy cutoff ($v \approx v_c$).
Note that the standard Kappa distribution with $\kappa=1$ is undefined,
because its energy density goes to infinity for $\kappa \le 3/2$.
However, we can still calculate
the density of the standard Kappa distribution for $\kappa > 1/2$.

We propose two rejection methods
to generate the regularized Kappa distribution in PIC simulation.
We first introduce a post-rejection method.
By setting $N_0=1$, we rewrite Eq.~\eqref{eq:RKD} in the following way,
\begin{align}
f_\mathrm{rk}(\vec{v}; \kappa, \theta, \alpha)
d^3v
=&
\left(
\dfrac{\Gamma(\kappa-1/2)}{\mathcal{U}\left( \frac{3}{2}, \frac{3}{2}-\kappa,\alpha^2\kappa\right) \Gamma(\kappa+1)}
\right)
\times
\bigg\{
f_{\kappa}(\vec{v}; \kappa, \theta, \theta)
\bigg\}
\times
\exp\left( -\frac{\alpha^2 v^2}{\theta^2} \right)
d^3{v}
\label{eq:RKD_decomposition}
\end{align}
In the right hand side,
the first bracketed term is a constant,
the second curly-bracketed term stands for the Kappa distribution (Eq.~\eqref{eq:kappa}), and
the last exponential term ranges from zero to unity. 
So, we generate the Kappa distribution first. 
The distribution is equivalent to
a multivariate $t$-distribution (Eq.~\eqref{eq:multi_t}),
i.e.,
\begin{align}
f_{\kappa}(\vec{v}; \kappa, \theta, \theta) d^3v
=
\mathrm{St}\left(\vec{v}; 2\kappa-1, \sqrt{\frac{\kappa\theta^2}{2\kappa-1}}, 3 \right) d^3v
\end{align}
and then we can use a Monte Carlo method,
prescribed by Eq.~\eqref{eq:t_gen} \citep{abdul15,zeni22}. 
After generating the Kappa distribution,
we further apply the rejection method.
We take the result, if a uniform variate $U \sim U(0,1)$ satisfy
\begin{align}
U < \exp\left({-\frac{\alpha^2 v^2}{\theta^2}}\right),
\label{eq:RKD_rej}
\end{align}
If this is not met, we discard the result.
The final velocity distribution follows Eq.~\eqref{eq:RKD}. 
The entire procedure is summarized in Algorithm 4.1 in Table \ref{table:RKD}.
The acceptance efficiency is given by an reciprocal of the bracketed term in
Eq.~\eqref{eq:RKD_decomposition}.
\begin{align}
{\rm eff}_{\rm rk}(\kappa, \alpha) =
\dfrac{\mathcal{U}\left( \frac{3}{2}, \frac{3}{2}-\kappa,\alpha^2\kappa\right) \Gamma(\kappa+1)}{\Gamma(\kappa-1/2)}
\label{eq:RKD_eff}
\end{align}
When $\alpha \rightarrow 0$, the rejection is rarely used, and the efficiency approaches ${\rm eff}_{\rm rk}(\kappa, 0) \rightarrow 1$.

\begin{table}[tpb]
\centering
\caption{Algorithms for the regularized Kappa distribution
\label{table:RKD}}
\begin{tabular}{l}
\hline
{\bf Algorithm 4.1: Post-rejection method} ($\kappa > 1/2$)\\
\hline
{\bf repeat}\\
~~~~generate $N_1, N_2, N_3 \sim \mathcal{N}(0,1)$ \\
~~~~generate $Y \sim \mathrm{Ga}(\kappa-1/2,2)$ \\
~~~~generate $U \sim U(0,1)$ \\
~~~~$v_x  \leftarrow \theta \sqrt{ \kappa } N_1 / \sqrt{Y}$ \\
~~~~$v_y  \leftarrow \theta \sqrt{ \kappa } N_2 / \sqrt{Y}$ \\
~~~~$v_z  \leftarrow \theta \sqrt{ \kappa } N_3 / \sqrt{Y}$ \\
{\bf until}~$U < \exp[ -\alpha^2 (v_x^2 + v_y^2 + v_z^2) /\theta^2 ]$\\
{\bf return} $v_x, v_y, v_z$\\
\hline
{\bf Algorithm 4.2: Piecewise rejection method} ($\kappa > 0$)\\
\hline
$x_c \leftarrow 1/(\alpha^2\kappa)$,\\
$S_L \leftarrow \frac{2}{1-2\kappa} \Big( (1+x_c)^{1/2-\kappa} - 1 \Big) ~~~~~(\mathrm{\kappa \ne 1/2})$ \\
$S_L \leftarrow \log ( 1 + x_c ) ~~~~~~~~~~~~~~~~~~~~~~~~(\mathrm{\kappa = 1/2})$ \\
$S_R = ({x_c^{3/2}}/{e}) \left(1 + x_c \right)^{-(\kappa+1)}$, \\
$p_L \leftarrow \dfrac{S_L}{S_L+S_R}$,~
$p_R \leftarrow \dfrac{S_R}{S_L+S_R}$\\
{\bf repeat}\\
~~~~generate $U_1, U_2 \sim U(0, 1)$ \\
~~~~{\bf if}~$U_1 \le p_L$ {\bf then} \\
~~~~~~~~$u \leftarrow {U_1}/{p_L}$ \\
~~~~~~~~$x \leftarrow [ u (1+x_c)^{1/2-\kappa} + (1-u) ]^{\frac{1}{1/2-\kappa}}-1$~~~\textrm{($\kappa \ne 1/2$)}\\
~~~~~~~~$x \leftarrow (1+x_c)^{u} - 1$~~~~~~~~~~~~~~~~~~~~~~~~~~~~~~~~~\textrm{($\kappa = 1/2$)}\\
~~~~~~~~{\bf if}~$U_2 < (x/(1+x))^{1/2} \exp\left({-x/x_c}\right)$ {\bf break} \\
~~~~{\bf else}  \\
~~~~~~~~$u \leftarrow (U_1-p_L)/{p_R} $ \\
~~~~~~~~$x \leftarrow x_c ( 1 - \log u )$ \\
~~~~~~~~{\bf if}~$U_2 < (x/x_c)^{1/2} \left( \frac{1 + x_c}{1 + x} \right)^{(\kappa+1)}$ {\bf break} \\
~~~~{\bf endif}\\
{\bf end repeat}
\\
generate $U_3, U_4 \sim U(0, 1)$ \\
$v \leftarrow \sqrt{ \kappa x } \theta$ \\
$v_x \leftarrow v ~ ( 2 U_3 - 1 )$ \\
$v_y \leftarrow 2 v \sqrt{ U_3 (1-U_3) } \cos(2\pi U_4)$ \\
$v_z \leftarrow 2 v \sqrt{ U_3 (1-U_3) } \sin(2\pi U_4)$ \\
{\bf return} $v_x, v_y, v_z$ \\
\hline
\end{tabular}
\end{table}

Next, we propose another procedure,
the piecewise rejection method.
Again we set $N_0=1$.
Using $x=v^2/(\kappa\theta^2)$,
we consider a distribution of $x$, i.e., $g(x)$.
\begin{align}
\label{eq:RKDx}
g(x) &\equiv
f_\mathrm{rk}({v}) 4\pi v^2
\left| \frac{dv}{dx} \right|
=
c_\mathrm{rk} \cdot
x^{1/2}
\left(1 + x\right)^{-(\kappa+1)}
\exp\left( -\alpha^2\kappa x\right)
\\
c_\mathrm{rk} &=
\frac{2}{\sqrt{\pi}~{\mathcal{U}\left(\frac{3}{2},\frac{3}{2}-\kappa, \alpha^{2}\kappa\right)}}
\end{align}
Here, $c_\mathrm{rk}$ is a constant.
We will develop a Monte Carlo procedure for $g(x)$.
We consider the following envelope functions,
\begin{align}
G(x)
&=
\left\{
\begin{array}{ll}
G_L(x) = c_\mathrm{rk}~
\left(1+x \right)^{-\kappa-1/2}
~~
& (0 \le x < x_c)
\\ [8pt]
G_R(x) 
=
c_\mathrm{rk}~
x_c^{1/2}
\left(1 + x_c \right)^{-(\kappa+1)}
\exp\left( - x /x_{c} \right)
~~~
& (x_c \le x < \infty)
\end{array}
\right.
\label{eq:RKD_envelope}
\end{align}
where $x_c = 1/(\alpha^2\kappa)$ is a cutoff value.
One can see $g(x) < G_L(x)$ for $0 \le x$.
In addition, since $\alpha < 1$,
the following $x$-derivative is negative for $(1/\kappa) < x_c \le x$,
\begin{align}
\frac{\partial}{\partial x}
\Big( x^{1/2} (1 + x)^{-(\kappa+1)} \Big)
=
\frac{1-\kappa x}{2 x^{1/2} (1+x)^{\kappa+2}}
< 0
\end{align}
This ensures $g(x) < G_R(x)$ for $x_c \le x$.
The areas below the envelopes are given by
\begin{align}
S_L &= \int_0^{x_c} G_L(x) dx =
\left\{
\begin{array}{ll}
c_\mathrm{rk}~
\frac{2}{1-2\kappa} \Big( (1+x_c)^{1/2-\kappa} - 1 \Big)
~~~
& (\kappa \ne 1/2) 
\\ [8pt]
c_\mathrm{rk}~
\log ( 1 + x_c )
~~~
& (\kappa = 1/2) 
\end{array}
\right.
\label{eq:RKD_SL}
\\
S_R &= \int_{x_c}^{\infty} G_R(x) dx =
c_\mathrm{rk}~
\frac{1}{e}
x_c^{3/2}
\left(1 + x_c \right)^{-(\kappa+1)}
\label{eq:RKD_SR}
\end{align}
where $e=2.718\dots$ is the Euler number.
Examining integral forms of the envelope functions,
we can construct inverse transform procedures to generate
the envelope distributions.
For example, when $\kappa \ne 1/2$, for the left area $S_L$,
we equate a uniform variate $u \sim U(0,1)$
\begin{align}
\Big( (1+x)^{1/2-\kappa} - 1 \Big)
=
u
\Big( (1+x_c)^{1/2-\kappa} - 1 \Big)
\end{align}
and then we find the following inverse transform procedure
\begin{align}
x \leftarrow \big[ u (1+x_c)^{1/2-\kappa} + (1-u) \big]^{\frac{1}{1/2-\kappa}}-1
.
\end{align}
Then, we construct rejection conditions,
by comparing the distribution (Eq.~\eqref{eq:RKDx}) and 
the envelope functions (Eq.~\eqref{eq:RKD_envelope}).
After obtaining $x$, we recover $v$ from $x$.
Then we randomly scatter $v$ into three directions.
The entire procedure is shown in Algorithm 4.2 in Table \ref{table:RKD}.
The procedure does not contain the coefficient $c_\mathrm{rk}$,
because it is cancelled away. 
The acceptance efficiency is given by
the reciprocal of the total area.
\begin{align}
{\rm eff}_{\rm rk2}(\kappa, \alpha)
&=
\frac{1}{ S_L + S_R }
=
\frac{\sqrt{\pi}~{\mathcal{U}\left(\frac{3}{2},\frac{3}{2}-\kappa, \alpha^{2}\kappa\right)} }{ 2 \left(
\frac{2}{1-2\kappa} \Big( (1+x_c)^{1/2-\kappa} - 1 \Big)
+
\frac{1}{e}
x_c^{3/2}
\left(1 + x_c \right)^{-(\kappa+1)}
\right)}
\label{eq:RKD_eff2}
\end{align}

We have carried out Monte Carlo tests of
our procedures in Table \ref{table:RKD}.
Using $10^6$ particles, we have generated
the regularized Kappa distribution ($(\kappa,\theta,\alpha)=(1, 1, 0.05)$) by the piecewise rejection method (Algorithm 4.2),
and then the particle distribution is shown by the blue histogram in Figure \ref{fig:RKD}(a).
The distribution is in excellent agreement with
the exact solution (the red line).
We have also generated
the standard Kappa distribution ($(\kappa,\theta,\alpha)=(1, 1, 0)$)
by the post-rejection method (Algorithm 4.1), and
their distribution is shown by the gray histogram.
It agrees with the exact solution (the black line).

Figure \ref{fig:RKD}(b) shows the acceptance efficiency of the two methods.
The solid and dashed lines present
theoretical predictions
for the post-rejection method (Eq.~\eqref{eq:RKD_eff}) and
for the piecewise rejection method (Eq.~\eqref{eq:RKD_eff2}), respectively.
Several $\alpha$ values are indicated by color.
The squares and the circles are numerical results,
in agreement with the predictions.

The post-rejection method can be used for $\kappa > 1/2$,
because it relies on the standard Kappa distribution.
At $\kappa$ approaches $1/2$,
the density in the power-law tail of the standard Kappa distribution
goes to infinity. 
The density beyond the cutoff velocity diverges, and then
the acceptance efficiency drops to zero. 
One may recall that
the standard Kappa distribution is defined for $\kappa > 3/2$,
where the second velocity moment remains finite.
For $1/2 < \kappa \le 3/2$,
the energy of the standard Kappa distribution diverges,
but we can still compute its density. 
Then, by applying the rejection method,
both the density and energy density of
the regularized Kappa distribution remain finite.
As $\kappa$ increases, the efficiency becomes better.
This is because
the power-law tail becomes less pronounced for large $\kappa$,
and because there are fewer particles beyond the cutoff velocity. 
By definition, the efficiency of the post-rejection method
approaches unity in the $\alpha \rightarrow 0$ limit.
For $\alpha=0.01$ (the black line and the black squares),
the acceptance efficiency is almost unity,
because of a very high cutoff velocity ($v_c = 100\theta$).
For $\alpha=0.1$ (the red line and the red squares), or
when the cutoff velocity is $v \gtrsim v_c = 10\theta$,
the acceptance efficiency is greater than 95\% for $\kappa > 2$.
As $\alpha$ increases, the algorithm becomes increasingly inefficient.
The $\alpha \gtrsim 1$ regime is unimportant, and out of the scope of this paper. 
Technically, as shown in Table \ref{table:RKD},
the post-rejection method is simple, and easy to implement.
The method requires caution when $1/2 < \kappa \le 3/2$. 
As described in Section \ref{sec:t} and in Algorithm 4.1,
the Kappa generator (the $t$-generator) uses one gamma variate.
When $1/2 < \kappa \le 3/2$, it has
the scale parameter less than unity, i.e., $\kappa-1/2 \le 1$.
In such a case, the gamma distribution ($\mathrm{Ga}(x)$) has non-zero density at $x=0$.
This may cause a zero-division problem in the algorithm.

The piecewise rejection method works in the entire range of $\kappa \ge 0$.
When $\kappa=1/2$, we need to switch some procedures, but
they are fully described in Algorithm 4.2 in Table \ref{table:RKD}. 
Its efficiency is moderate.
It is not unity in the $\alpha \rightarrow 0$ limit.
When $\alpha \rightarrow +0$ and $\kappa > 1/2$,
from Eqs.~\eqref{eq:RKDx} and \eqref{eq:RKD_envelope}, we find
\begin{align}
{\rm eff}_{\rm rk2}(\kappa, 0)
&\approx
\dfrac
{ \int_0^\infty x^{1/2} (1+x)^{-\kappa-1} dx }
{ \int_0^\infty (1+x)^{-\kappa-1/2} dx }
=
\dfrac
{
\frac{\sqrt{\pi} \Gamma(\kappa-\frac{1}{2})}{ 2 \Gamma(\kappa+1) }
\int_0^\infty B' \left( x; \frac{3}{2}, \kappa-\frac{1}{2}, 1, 1 \right) dx }
{ 1 / ( \kappa - 1/2 ) }
=
\frac{\sqrt{\pi} \Gamma(\kappa+\frac{1}{2})}{ 2 \Gamma(\kappa+1) }
\label{eq:RKD_eff3}
\end{align}
where we utilized the beta-prime distribution (Eq.~\eqref{eq:gbetap}).
When $\alpha \rightarrow +0$ and $0 < \kappa < 1/2$,
Eq.~\eqref{eq:RKD_eff2} can be approximated to
\begin{align}
{\rm eff}_{\rm rk2}(\kappa, 0)
&\approx
\frac{\sqrt{\pi}~{\mathcal{U}\left(\frac{3}{2},\frac{3}{2}-\kappa, \frac{1}{x_c} \right)} }{ 2 \left(
\frac{2}{1-2\kappa}
+
\frac{1}{e}
\right)
x_c^{1/2-\kappa}
}
\approx
\frac{ \Gamma(1/2-\kappa) }{ \left(
\frac{2}{1-2\kappa}
+
\frac{1}{e}
\right)
}
\label{eq:RKD_eff4}
\end{align}
with help from the asymptotic relation for $x_c\rightarrow \infty$
\citep[][Section 13, 13.2.18]{dlmf},
\begin{align}
{\mathcal{U}\left(\frac{3}{2},\frac{3}{2}-\kappa, \frac{1}{x_c} \right)}
&\approx
\frac{\Gamma(1/2-\kappa)}{\Gamma(3/2)}
{x_c}^{1/2-\kappa}
+
\frac{\Gamma(\kappa-1/2)}{\Gamma(\kappa+1)}
+
\mathcal{O}\left( x_c^{-1/2-\kappa} \right)
\end{align}
In Figure \ref{fig:RKD}(b),
the purple dot-dashed line and dotted line indicate
estimates of the $\alpha \rightarrow 0$ limit (Eqs.~\eqref{eq:RKD_eff3} and \eqref{eq:RKD_eff4}).
They give good upper bounds to the efficiency.
As $\alpha$ increases, the piecewise rejection method becomes less efficient,
similarly as the post-rejection method. 
In the $\kappa \rightarrow +0$ limit,
the method gives similar results regardless of $\alpha$,
because the distribution (Eq.~\eqref{eq:RKDx}) resembles
a Gamma distribution, which is scale-free.
\begin{align}
g(x) \approx
c_\mathrm{rk} \cdot
x^{-\kappa-1/2}
\exp\left( - x/x_c \right)
\propto
\mathrm{Ga}(x; 1/2-\kappa, x_c )
\end{align}
The efficiencies converge to
$\mathrm{eff}_\mathrm{rk2}(0,0) \approx \sqrt{\pi} / (2+1/e) \approx 0.7485$
as estimated by Eq.~\eqref{eq:RKD_eff4}.

In summary, we have discussed the two methods.
The piecewise rejection method is the only choice
for $0<\kappa \le 1/2$.
Either method can be used for $1/2 < \kappa \le 3/2$, but
caution is needed for the post-rejection method.
For $\kappa > 3/2$, we recommend the post-rejection method,
which outperforms the piecewise rejection method.

\section{Subtracted Kappa distribution}
\label{sec:SK}

To date, several loss-cone distribution models have been developed.
To control the loss-cone angle,
some models use a power of the perpendicular velocity,
i.e., $(v_\perp)^{2j}$ \citep{DGH}, and
others use a power of the pitch-angle sine,
i.e., $(\sin \alpha)^{2j}=(v_\perp/v)^{2j}$ \citep{kennel66},
where $j\ge 0$ is the loss-cone index and $\alpha$ is the pitch angle.
These models are not easy to use,
because a narrow loss cone in the Earth's magnetosphere
requires the index less than unity. 
Instead, the subtracted Maxwellian distribution \citep{AK78} has been widely used.
It has intermediate characteristics between the \citet{DGH} models with $j=0$ and $j=1$.

The Kappa loss-cone (KLC) distributions have been studied
for decades \citep{summers91,xiao98}.
Similarly, the loss-cone is approximated by
a power of the perpendicular velocity \citep{summers91}
or the pitch-angle sine \citep{xiao98,zeni23}, but
it was very recently that
\citet{summers25} proposed the subtracted Kappa distribution,
a Kappa extension of the subtracted Maxwellian.
This new model is promising for studying kinetic processes
in the KLC distribution with a narrow loss-cone.
The phase space density $f_\mathrm{sk}$, its normalization constant $C_\mathrm{sk}$, pressures, energy density, and the pressure anisotroy $A$ are given by:
\begin{align}
\label{eq:SK}
f_\mathrm{sk}  (v_{\parallel}, \vec{v}_\perp)
\,d^3 v
&=
N_0 \cdot C_\mathrm{sk}
\left\{
\frac{1-\Delta\beta}{1-\beta}
\left( 1 + \frac{ {v_{\parallel}}^2 }{\kappa \theta_{\parallel}^2}  + \frac{ v_{\perp}^2 }{\kappa \theta_{\perp}^2} \right)^{-(\kappa+1)}
-
\frac{1-\Delta}{1-\beta}
\left( 1 + \frac{ {v_{\parallel}}^2 }{\kappa \theta_{\parallel}^2}  + \frac{ v_{\perp}^2 }{\beta \kappa \theta_{\perp}^2} \right)^{-(\kappa+1)}
\right\}
\,d^3 v
\\
&=
N_0 \cdot C_\mathrm{sk}
\Bigg\{
\Delta\cdot
\left( 1 + \frac{ {v_{\parallel}}^2 }{\kappa \theta_{\parallel}^2}  + \frac{ v_{\perp}^2 }{\kappa \theta_{\perp}^2} \right)^{-(\kappa+1)}
\nonumber \\
& ~~~~~~~~~~~~~~~~~~~~
+
\frac{1-\Delta}{1-\beta}
\left[
\left( 1 + \frac{ {v_{\parallel}}^2 }{\kappa \theta_{\parallel}^2}  + \frac{ v_{\perp}^2 }{\kappa \theta_{\perp}^2} \right)^{-(\kappa+1)}
-
\left( 1 + \frac{ {v_{\parallel}}^2 }{\kappa \theta_{\parallel}^2}  + \frac{ v_{\perp}^2 }{\beta \kappa \theta_{\perp}^2} \right)^{-(\kappa+1)}
\right]
\Bigg\}
\,d^3 v
\label{eq:SK_weight}
\\
C_\mathrm{sk} &=
\frac{\Gamma(\kappa+1)}{(\kappa\pi)^{3/2}\theta_{\parallel}\theta_{\perp}^2 \Gamma(\kappa-\frac{1}{2})},
\\
P_{\parallel} &= \frac{\kappa}{2\kappa-3} N_0 m \theta_{\parallel}^2,
~~
P_{\perp} = \frac{\kappa ( 1 + \beta [ 1 -\Delta] )}{2\kappa-3} N_0 m \theta_{\perp}^2,
\\
\mathcal{E}
&=
\frac{\kappa}{2\kappa-3} N_0 m
\left(
\frac{1}{2} \theta_{\parallel}^2
+ ( 1 + \beta [ 1 -\Delta] ) \theta_{\perp}^2
\right),
\\
A &\equiv
\frac{P_{\perp}}{P_{\parallel}} - 1 =
( 1 + \beta [ 1 -\Delta] )
\frac{ \theta_{\perp}^2 }{ \theta_{\parallel}^2 }
- 1
\end{align}
Here, the kappa index $\kappa > 3/2$ controls a slope in the high-energy tail,
the shape parameter $\beta \in [0,1]$ controls the opening angle of the loss cone,
the filling parameter $\Delta$ controls the density inside the loss cone,
and
$\theta_{\parallel}$ and $\theta_{\perp}$ are the parallel and perpendicular characteristic velocities.
When $\kappa \rightarrow \infty$, the distribution reverts to
the subtracted Maxwellian. 
When $\beta=0$ or $\Delta=1$, it becomes a bi-Kappa distribution with no loss cone.
When $\beta=1$, the distribution approaches
the Dory-type (Summers-type) KLC distribution
with loss-cone index $j=1$ \citep[][$\sigma=1$ in their notation]{summers91}. 
When $\Delta=0$, the loss cone is well developed.
When $\Delta=1$, the loss cone is completely filled, and then
the distribution reverts to a bi-Maxwellian.

\begin{table}
\centering
\caption{Algorithm for the subtracted Kappa distribution
\label{table:SK}}
\begin{tabular}{l}
\\
\hline
{\bf Algorithm 5: subtracted Kappa} ($\kappa>3/2$)\\
\hline
generate $U_1, U_2, U_3 \sim U(0,1)$ \\
generate $N \sim \mathcal{N}(0,1)$ \\
generate $Y \sim \mathrm{Ga}(\kappa-1/2,2)$ \\
$x \leftarrow -\log U_1 - \beta \log \left( \min\left( \frac{U_2}{1-\Delta},1\right) \right)$ \\
$v_{\perp 1}  \leftarrow \theta_{\perp} \sqrt{ 2\kappa x} \cos(2\pi U_3) / \sqrt{Y}$\\
$v_{\perp 2}  \leftarrow \theta_{\perp} \sqrt{ 2\kappa x} \sin(2\pi U_3) / \sqrt{Y}$\\
$v_{\parallel} ~ \leftarrow \theta_{\parallel} \sqrt{ \kappa } N / \sqrt{Y}$ \\
{\bf return} $v_{\perp 1}, v_{\perp 2}, v_{\parallel}$\\
\hline
\end{tabular}
\end{table}

Since it was proposed very recently \citep{summers25},
there is no Monte Carlo algorithm for the distribution.
Here, we propose a compact procedure
for randomly generating the subtracted Kappa distribution
in Algorithm 5 (Table \ref{table:SK}).
Below, we explain why the procedure gives a random variate
that is distributed by Eq.~\eqref{eq:SK}.

For a moment, we assume $\Delta=0$.
We combine two uniform variates $U_1 $ and $U_2$ in the following way,
\begin{align}
x \leftarrow -\log U_1 - \beta \log U_2
\label{eq:subM_x}
\end{align}
The two terms correspond to the exponential distributions (Section \ref{sec:exp}).
The probability distribution of $x$ ($G(x)$) is given by
their convolution \citep[][Section 2]{zeni23}.
\begin{align}
G(x)
&=
\int_0^x
\bigg\{
\mathrm{Exp}(s; 1)
\times
\mathrm{Exp}(x-s; \beta)
\bigg\}
\,
ds
\\
&=
\frac{1}{1-\beta}
\left\{
\exp \left( - x \right)
-
\exp \left( - \frac{x}{\beta} \right)
\right\}
=
\frac{1}{1-\beta}
e^{-x}
-
\frac{\beta}{1-\beta}
\left(
\frac{1}{\beta}
e^{- x/{\beta}}
\right)
\label{eq:subM}
\end{align}
Eq.~\eqref{eq:subM} indicates that
$G(x)$ is equivalent to the subtraction of two exponential distributions.

Next we consider three independent variates of $E$, $N$, and $Y$.
The first variate $E$ follows the exponential distribution with scale $\lambda$
(i.e., $E \sim \mathrm{Exp}(\lambda)$),
$N$ follows the standard normal distribution
(i.e., $N \sim \frac{1}{\sqrt{2\pi}} e^{-x^2/2}$), and
$Y$ follows a gamma distribution with shape $\kappa-1/2$ and scale $2$
(i.e., $Y \sim \mathrm{Ga}(\kappa-1/2,2)$).
Note that we retain the scale parameter $\lambda$
in the exponential distribution and subsequent equations.
Using them, we consider the following variables.
\begin{align}
\label{eq:V1}
V_{\perp}^2 &= 2\theta_{\perp}^2 \kappa E  / {Y} \\
\label{eq:V2}
V_{\parallel} &= \theta_{\parallel} \sqrt{ \kappa } N / \sqrt{Y} \\
Z &= \frac{2}{\lambda} E + N^2 + Y
\label{eq:Z}
\end{align}
They yield the following relation and the Jacobian
\begin{align}
\label{eq:E}
E & 
= \frac{V_{\perp}^2}{2\kappa\theta_{\perp}^2} Z \left( 1 + \frac{V_{\parallel}^2}{\kappa\theta_{\parallel}^2} + \frac{V_{\perp}^2}{\lambda \kappa\theta_{\perp}^2} \right)^{-1} \\
\label{eq:N}
N &
= \frac{V_{\parallel}}{\sqrt{\kappa}\theta_{\parallel}} \sqrt{Z} 
\left( 1 + \frac{V_{\parallel}^2}{\kappa\theta_{\parallel}^2} + \frac{V_{\perp}^2}{\lambda \kappa\theta_{\perp}^2} \right)^{-1/2} \\
\label{eq:Y}
Y &= Z \left( 1 + \frac{V_{\parallel}^2}{\kappa\theta_{\parallel}^2} + \frac{V_{\perp}^2}{\lambda \kappa\theta_{\perp}^2} \right)^{-1}
\\
\left\| \frac{\partial(E, N, Y)}{\partial(V_{\perp}, V_{\parallel}, Z)} \right\|
&=
\left(~
\left\| \frac{\partial(V_{\perp}, V_{\parallel}, Z)}{\partial(E, N, Y)} \right\|
~\right)^{-1}
=
\frac{V_{\perp}Z^{3/2}}
{ \kappa^{3/2} \theta_\parallel \theta_\perp^2 }
\left( 1 + \frac{V_{\parallel}^2}{\kappa\theta_{\parallel}^2} + \frac{V_{\perp}^2}{\lambda \kappa\theta_{\perp}^2} \right)^{-5/2}
\label{eq:Jacobian}
\end{align}
Then we consider the joint probability distribution function of $E$, $N$, and $Y$.
It is given by a product of the exponential distribution of $\varepsilon$, the normal distribution of $n$, and the gamma distribution of $y$,
because these three distributions are independent.
\begin{align}
f_{E,N,Y}(\varepsilon,n,y) = 
\left(
\frac{1}{w} e^{-\varepsilon/w}
\right)
\times
\left(\frac{1}{\sqrt{2\pi}}e^{-n^2/2} \right)
\times
\left( \frac{ y^{\kappa-3/2} e^{ -y / 2 } }{ \Gamma(\kappa-1/2) 2^{\kappa-1/2}} \right)
\label{eq:fXNY}
\end{align}
Using Eqs.~\eqref{eq:E}--\eqref{eq:Jacobian},
we rewrite the joint distribution
with respect to $V_{\perp}$, $V_{\parallel}$, and $Z$.
\begin{align}
f_{V_{\perp},V_{\parallel},Z}(v_{\perp},v_{\parallel},z)
&= 
\frac{1}{w\sqrt{2\pi} \Gamma(\kappa-1/2) 2^{\kappa-1/2}}
\left(
e^{-\varepsilon/w}
\right)
\times
\left(e^{-n^2/2} \right)
\times
\left( { y^{\kappa-3/2} e^{ -y / 2 } } \right)
\left\| \frac{\partial(E, N, Y)}{\partial(V_{\perp}, V_{\parallel}, Z)} \right\|
\nonumber \\
&= 
\left(
\frac{ z^{\kappa} e^{-z/2} }{ \Gamma(\kappa+1) 2^{\kappa+1}}
\right)
\times
\left(
2\pi v_{\perp}
\frac{ \Gamma(\kappa+1) }
{\lambda (\pi\kappa)^{3/2} \theta_\parallel \theta_\perp^2 \Gamma(\kappa-1/2) }
\left[ 1 + \frac{v_{\parallel}^2}{\kappa\theta_{\parallel}^2} + \frac{v_{\perp}^2}{\lambda \kappa\theta_{\perp}^2} \right]^{-(\kappa+1)}
\right)
\label{eq:proof}
\end{align}
Eq.~\eqref{eq:proof} tells us that
$Z$ is distributed by the gamma distribution with shape $\kappa+1$ and scale $2$,
i.e., $Z \sim \mathrm{Ga}(\kappa+1,2)$,
and that the other variables $V_{\perp}$ and $V_{\parallel}$ are distributed
by a bi-Kappa distribution (an anisotropic Kappa distribution). 
These two, the gamma and the bi-Kappa distributions, are independent. 
In short, a procedure by Eqs.~\eqref{eq:V1}--\eqref{eq:Z}
gives the bi-Kappa distribution.

From Eqs.~\eqref{eq:subM} and \eqref{eq:proof},
we see that the procedure of Eqs.~\eqref{eq:subM_x}, \eqref{eq:E}--\eqref{eq:Y} gives the following distribution
\begin{align}
&
\frac{1}{1-\beta}
\left(
2\pi v_{\perp}
C_\mathrm{sk}
\left[ 1 + \frac{v_{\parallel}^2}{\kappa\theta_{\parallel}^2} + \frac{v_{\perp}^2}{\kappa\theta_{\perp}^2} \right]^{-(\kappa+1)}
\right)
-
\frac{\beta}{1-\beta}
\left(
2\pi v_{\perp}
\frac{ C_\mathrm{sk} }{\beta}
\left[ 1 + \frac{v_{\parallel}^2}{\kappa\theta_{\parallel}^2} + \frac{v_{\perp}^2}{\beta \kappa\theta_{\perp}^2} \right]^{-(\kappa+1)}
\right)
\\
&
~~~~~~~~~~~~
=
\frac{
2\pi v_{\perp}
C_\mathrm{sk}
}{1-\beta}
\left\{
\left( 1 + \frac{v_{\parallel}^2}{\kappa\theta_{\parallel}^2} + \frac{v_{\perp}^2}{\kappa\theta_{\perp}^2} \right)^{-(\kappa+1)}
-
\left( 1 + \frac{v_{\parallel}^2}{\kappa\theta_{\parallel}^2} + \frac{v_{\perp}^2}{\beta \kappa\theta_{\perp}^2} \right)^{-(\kappa+1)}
\right\}
\end{align}
We obtain the $\Delta=0$ case
by translating $v_\perp \rightarrow (v_{\perp 1}, v_{\perp 2})$.
Thus, the procedure gives the subtracted Kappa distribution when $\Delta=0$.

Finally, we discuss the $\Delta \ne 0$ cases.
Like the subtracted Maxwellian,
the subtracted Kappa distribution is designed as
a weighted sum of
the standard Kappa distribution ($\Delta$) and
the subtracted parts ($1-\Delta$),
as shown in Eq.~\eqref{eq:SK_weight}. 
We can generate the standard Kappa part
by setting $\beta$ to be zero.
To generate a full distribution of Eq.~\eqref{eq:SK},
we replace Eq.~\eqref{eq:subM_x} by the following procedure,
to disable the $\beta$ term at the probability of $\Delta$.
\begin{align}
x \leftarrow -\log U_1 - \beta \log \left( \min\left( \frac{U_2}{1-\Delta},1\right) \right)
\label{eq:subM_x2}
\end{align}
A full recipe consists of Eqs.~\eqref{eq:subM_x2}, \eqref{eq:V1}, and \eqref{eq:V2},
as shown in Algorithm 5 (Table \ref{table:SK}).

\begin{figure*}[htbp]
\centering
\includegraphics[width={\textwidth}]{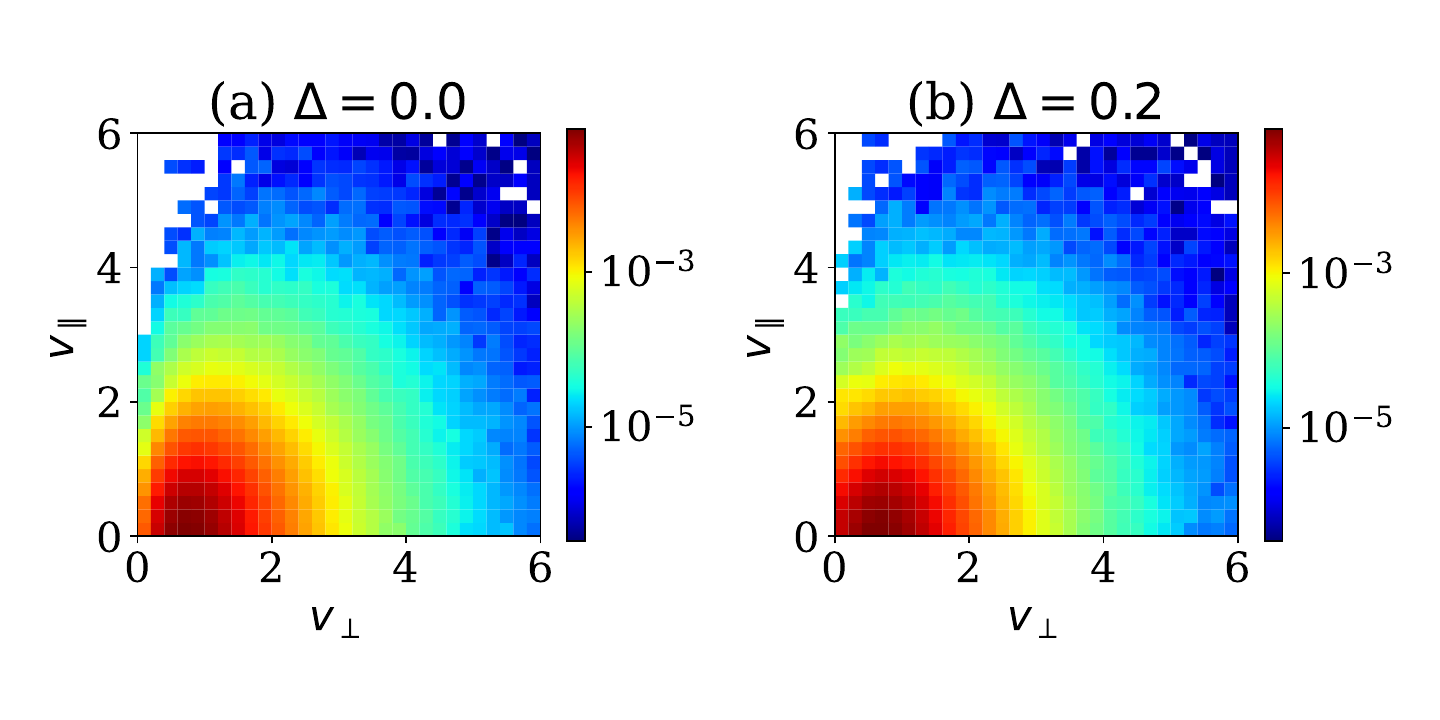}
\caption{
Monte Carlo sampling of the subtracted Kappa distribution ($\theta=1.0$, $\kappa=3.5$, $\beta=0.5$) with $10^6$ particles.
Phase space density is shown in the $v_{\perp}$--$v_{\parallel}$ plane.
The filling parameter is (a) $\Delta=0.0$ and (b) $\Delta=0.2$, respectively.
\label{fig:SK}}
\end{figure*}

We have verified our method
by a Monte Carlo test with $10^6$ particles.
Figure \ref{fig:SK} shows the phase space density of
the subtracted Kappa distribution ($\theta=1.0$, $\kappa=3.5$, $\beta=0.5$)
in the $v_{\perp}$--$v_{\parallel}$ plane.
The filling parameters are (a) $\Delta=0.0$ and (b) $\Delta=0.2$, respectively.
We recognize a narrow loss-cone in the $\Delta=0.0$ case and
a partially-filled loss-cone in the $\Delta=0.2$ case.
Although not shown here, 
we have compared the numerical results and the exact solution
at several 1-D cuts in the velocity space, and
found that they are in excellent agreement.

\section{Ring and Shell distributions}
\label{sec:ring}

In the solar wind, pickup ions (PUIs) lead to
a ring-shaped component in the velocity space.
They further evolve to an isotropic, shell-shaped distribution,
because of pitch-angle scattering \citep{vas76}.
Considering PUI's importance in kinetic processes and
their mathematical similarity,
we discuss ring and shell distributions together in this section. 

We first discuss a ring distribution with a Gaussian width
\citep[e.g.,][]{wu89,kumar97}.
We show the phase space density $f_\mathrm{ring}(\vec{v})$,
its perpendicular part $g_\mathrm{ring}(v_\perp)$, and
an auxiliary function $C_2(x)$.
\begin{align}
\label{eq:ring0}
f_\mathrm{ring}(\vec{v})d^3{v}
&=
N_0 \frac{1}{{\pi}^{3/2} \theta_\parallel \theta_\perp^2 C_2(\frac{V}{\theta_\perp})}
\exp\left( -\frac{v_\parallel^2}{\theta_\parallel^2} -\frac{(v_\perp-V)^2}{\theta_\perp^2}\right)
d^3{v}
,
\\
&=
N_0
\left\{
\frac{1}{\sqrt{\pi} \theta_\parallel}
\exp\left( -\frac{v_\parallel^2}{\theta_\parallel^2} \right)
dv_{\parallel}
\right\}
\times
\bigg\{
g_\mathrm{ring}({v}_\perp)
dv_{\perp}
\bigg\}
,
\\
\label{eq:ring_g}
g_\mathrm{ring}({v}_\perp)
&=
\frac{2 v_\perp}{\theta_\perp^2 C_2(\frac{V}{\theta_\perp})}
\exp\left(
-\frac{(v_\perp-V)^2}{\theta_\perp^2}\right),
\\
C_2(x) &= \exp( -x^2 ) + \sqrt{\pi} x\, \mathrm{erfc}\left( -x \right)
,
\end{align}
Here, $V$ is the ring velocity, $\theta_\parallel$ and $\theta_\perp$ are parallel and perpendicular characteristic velocities, and $\mathrm{erfc}(x)$ is the complementary error function.
The perpendicular and parallel components are independent.
The parallel component is assumed to be a Maxwellian. 
In the context of the solar wind,
$V$ corresponds to the solar-wind speed in the perpendicular direction. 
The pressures, the energy density, and the pressure anisotropy are given by \citep{kumar97}
\begin{align}
P_{\perp}
&= \frac{N_0 m \theta_\perp^2}{4 C_2(\frac{V}{\theta_\perp})}
\left\{
2\left( 1+ \frac{V^2}{\theta_\perp^2} \right)
\exp\left(-\frac{V^2}{\theta_\perp^2}\right)
+
\sqrt{\pi}
\frac{V}{\theta_\perp}
\left( 3 + 2 \frac{V^2}{\theta_\perp^2} \right)\,
\mathrm{erfc}\left( -\frac{V}{\theta_\perp} \right)
\right\} \\
&= \frac{N_0 m \theta_\perp^2}{2}
\left\{
1 + \frac{V^2}{\theta_\perp^2}
+
\frac{ \sqrt{\pi} }{2C_2(\frac{V}{\theta_\perp})}
\frac{V}{\theta_\perp}
\mathrm{erfc}\left( -\frac{V}{\theta_\perp} \right)
\right\} \\
P_{\parallel}
&= \frac{1}{2}N_0 m \theta_\parallel^2,~~~~
\mathcal{E} =
\frac{1}{4} N_0 m
\left\{
\theta_\parallel^2 +
{2\theta_\perp^2}
\left[
1 + \frac{V^2}{\theta_\perp^2}
+
\frac{ \sqrt{\pi} }{2C_2(\frac{V}{\theta_\perp})}
\frac{V}{\theta_\perp}
\mathrm{erfc}\left( -\frac{V}{\theta_\perp} \right)
\right]
\right\} 
\\
A &=
\frac{\theta_\perp^2}{\theta_\parallel^2}
\left\{
1 + \frac{V^2}{\theta_\perp^2}
+
\frac{ \sqrt{\pi} }{2C_2(\frac{V}{\theta_\perp})}
\frac{V}{\theta_\perp}
\mathrm{erfc}\left( -\frac{V}{\theta_\perp} \right)
\right\}
-1
.
\end{align}

We also discuss the shell distribution with a Gaussian width \citep[e.g.,][]{freund88,yoon89,liu11,min15}. 
We show the phase space density $f_\mathrm{shell}(\vec{v})$,
its radial form $g_\mathrm{shell}(v)$, and
an auxiliary function $C_3(x)$.
\begin{align}
\label{eq:shell0}
f_\mathrm{shell}(\vec{v})d^3{v}
&=
\frac{N_0}{2 \pi \theta^3 C_3(\frac{V}{\theta})}
\exp\left( -\frac{(|v|-V)^2}{\theta^2} \right)
d^3{v}
=
N_0 \bigg\{
g_\mathrm{shell}({v})
dv
\bigg\}
,
\\
g_\mathrm{shell}({v})
&=
\frac{2 v^2}{\theta^3 C_3(\frac{V}{\theta})}
\exp\left(
-\frac{(|v|-V)^2}{\theta^2}\right),
\label{eq:shell_g}
\\
C_3(x) &= x \exp({- x^2}) + \sqrt{\pi} \left(x^2 + \frac{1}{2} \right) \mathrm{erfc}{\left(-x \right)}
.
\end{align}
where $V$ is the shell velocity and $\theta$ is the characteristic velocity. 
We only consider the isotropic case. 
The pressure and the energy density are given by
\begin{align}
{P}
&= \frac{N_0 m \theta^2}{12 C_3(\frac{V}{\theta})}
\left\{
2 \left( 5 + 2\frac{V^2}{\theta^2} \right)
\frac{V}{\theta} \exp\left(-\frac{V^2}{\theta^2}\right)
+
\sqrt{\pi}
\left( 3+ 12\frac{V^2}{\theta^2} + 4\frac{V^4}{\theta^4} \right)
\mathrm{erfc}\left(-\frac{V}{\theta}\right)
\right\}
\\
&= \frac{N_0 m \theta^2}{6}
\left\{
5 + 2\frac{V^2}{\theta^2}
-
\frac{ \sqrt{\pi} }{ C_3(\frac{V}{\theta}) }
\mathrm{erfc}\left(-\frac{V}{\theta}\right)
\right\}
\\
\mathcal{E} &=
\frac{N_0 m \theta^2}{4}
\left\{
5 + 2\frac{V^2}{\theta^2}
-
\frac{ \sqrt{\pi} }{ C_3(\frac{V}{\theta}) }
\mathrm{erfc}\left(-\frac{V}{\theta}\right)
\right\}
\end{align}
When $V=0$, the distribution reverts to an isotropic Maxwellian. 
As we can imagine, both ring and shell distributions are
artificially truncated at $v_{\perp}=0$ or at $v=0$.
This won't be a big issue for $\theta \lesssim V$,
but becomes problematic for $V \lesssim \theta$.

To draw the ring and shell distributions in PIC simulation,
perhaps the inverse transform method is used,
even though it is not well documented. 
It is trivial to generate the parallel component of the ring distribution.
\citet{umeda07} inverted the perpendicular velocity
from the CDF of the perpendicular distribution (Eq.~\eqref{eq:ring_g}).
This can be implemented
by using a numerical table and a uniform variate $U \sim U(0,1)$.
\begin{align}
\mathrm{CDF}_\mathrm{ring}(v_\perp)
&= 1 - \frac{ \exp\left(-\frac{(v_\perp-V)^2}{\theta_\perp^2}\right)
+ \sqrt{\pi} \frac{V}{\theta_\perp}\, \mathrm{erfc}\left( \frac{v_\perp - V}{\theta_\perp} \right)
}
{C_2(\frac{V}{\theta})}
\label{eq:cdf2d}
\\
v_\perp &= \mathrm{CDF^{-1}_{ring}}(U)
\end{align}
For the shell distribution,
one can similarly construct the inverse transform method. 
\begin{align}
\mathrm{CDF}_\mathrm{shell}(v)
&= 1 - \frac{\frac{(v + V)}{\theta} \exp\left(-\frac{(v-V)^2}{\theta^2}\right)
+ \sqrt{\pi} \left(\frac{V^2}{\theta^2} + \frac{1}{2} \right) \mathrm{erfc}{\left(\frac{v-V}{\theta} \right)}
}
{C_3(\frac{V}{\theta})}
\label{eq:cdf3d}
\\
v &= \mathrm{CDF^{-1}_{shell}}(U)
\end{align}
Instead of the inverse transform, one can also use
a piecewise rejection method for log-concave distributions \citep{devroye86},
as described in Appendix A. 
Aside from detail implementation of the inverse transform,
the procedures for the ring and shell distributions are listed
in Algorithms 6.1 and 6.2 in Table \ref{table:ringshell}.

\begin{figure*}[htbp]
\centering
\includegraphics[width={\textwidth}]{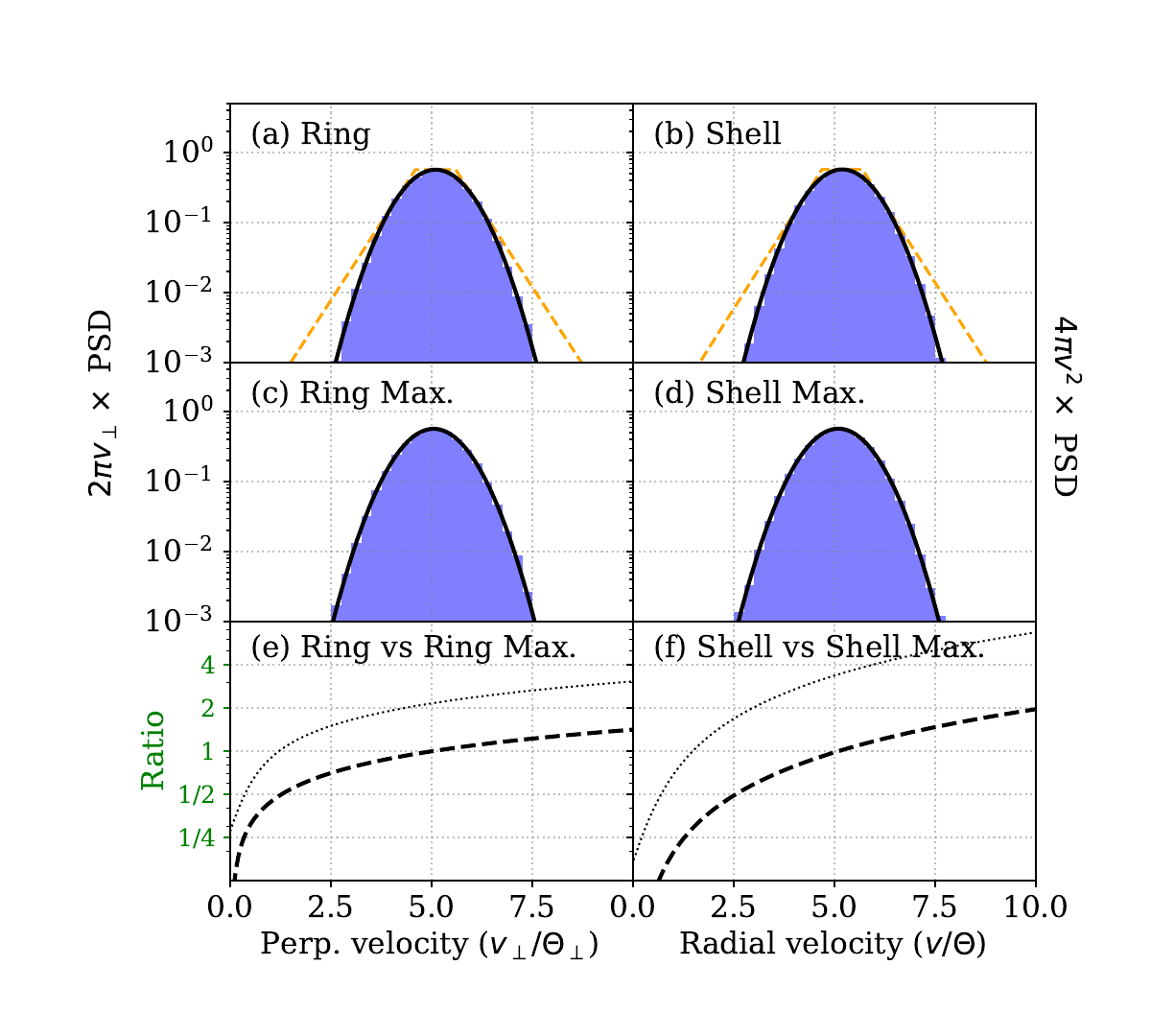}
\caption{
Left) Perpendicular velocity distributions of the ring distributions,
$2\pi v_\perp \times g_\textrm{ring}(v_\perp)$. 
Right) Radial velocity distributions of the shell distributions,
$4\pi v^2 \times f(v)$. 
(a) The ring distribution (Eq.~\eqref{eq:ring0}),
(b) the shell distribution (Eq.~\eqref{eq:shell0}),
(c) the ring Maxwellian (Eq.~\eqref{eq:ringM}),
and
(d) the shell Maxwellian (Eq.~\eqref{eq:shellM})
are presented by the black lines.
The blue histograms are numerical results with $10^6$ particles.
(e) The ratios of the ring distribution to the ring Maxwellian
for $V=5\theta_\perp$ (dashed lines) and for $V=\theta_\perp$ (dotted lines).
(f) The ratios of the shell distribution to the shell Maxwellian
for $V=5\theta$ (dashed lines) and for $V=\theta$ (dotted lines).
The orange dashed lines in Panels (a,b) are envelope functions for the piecewise rejection method, described in Appendix A.
\label{fig:shell}}
\end{figure*}

Figures~\ref{fig:shell}(a) and (b) shows
perpendicular and radial particle distributions
for $V=5\theta_\perp$ or $V=5\theta$, together with numerical results.
The blue histograms show Monre-Carlo results with $10^6$ particles.
They agree with the exact solutions (Eqs.~\eqref{eq:ring_g} and \eqref{eq:shell_g}),
indicated by the black lines, very well.

\begin{table}
\centering
\caption{Algorithms for ring and shell distributions.
For the inverse transform method, please see the text.
For the piecewise rejection method,
see Appendix A and the procedure in Table.~\ref{table:swisdak}.
\label{table:ringshell}}
\begin{tabular}{l}
\hline
{\bf Algorithm 6.1: ring distribution}\\
\hline
generate $U_1, U_2 \sim U(0, 1)$  \\
generate $N \sim \mathcal{N}(0,1)$ \\
$R \leftarrow \mathrm{CDF_{ring}^{-1}}(U_1)$ or $\mathrm{PiecewiseRejection}()$ \\
$v_{\perp 1} \leftarrow \theta_{\perp} R \cos(2\pi U_2)$ \\
$v_{\perp 2} \leftarrow \theta_{\perp} R \sin(2\pi U_2)$ \\
$v_{\parallel} ~ \leftarrow \theta_{\parallel} N / \sqrt{2}$ \\
{\bf return} $v_{\perp 1}, v_{\perp 2}, v_{\parallel}$\\
\hline
{\bf Algorithm 6.2: shell distribution}\\
\hline
generate $U_1, U_2 , U_3 \sim U(0, 1)$  \\
$R \leftarrow \mathrm{CDF^{-1}_{shell}}(U_1)$ or $\mathrm{PiecewiseRejection}()$ \\
$v_{x} \leftarrow ~ R ~ ( 2 U_2 - 1 )$ \\
$v_{y} \leftarrow 2 R \sqrt{ U_2 (1-U_2) } \cos(2\pi U_3)$ \\
$v_{z} \leftarrow 2 R \sqrt{ U_2 (1-U_2) } \sin(2\pi U_3)$ \\
{\bf return} $v_x, v_y, v_z$ \\
\hline
\end{tabular}
\end{table}

\section{Ring and Shell Maxwellians}
\label{sec:ringMax}

We discuss another ring and shell-shaped distributions,
based on the scattering of seed Maxwellians.
By gyrating a seed Maxwellian population,
\citet{usami22} has proposed the following axisymmetric distribution,
\begin{align}
f_\mathrm{rM}(\vec{v}) d^3v &=
\frac{N_0}{\pi^{3/2} \theta_\parallel\theta_\perp^2}
\exp\left(
-\frac{v_{\parallel}^2}{\theta_\parallel^2} 
-\frac{v_\perp^2+V^2}{\theta_\perp^2}
\right)
I_0\left(\frac{2 v_\perp V}{\theta_\perp^2}\right)
d^3v
\label{eq:ringM}
\end{align}
where $V$ is the ring velocity and
$I_0(x)$ is the zeroth-order modified Bessel function of the first kind.
We call this distribution the ring Maxwell distribution or the ring Maxwellian\footnote{The original authors call the distribution the ``pseudo Maxwellian,'' because they study near-Maxwellian cases, $V \lesssim \theta_\perp$. In this study, we call it the ring Maxwellian to emphasize its role as a ring-type distribution.}.
Note that we employ a different convention for the characteristic velocity from the original paper, by a factor of $\sqrt{2}$. 
The phase space density is concave at $v_\perp \simeq 0$ when $V \le \theta_\perp$,
while it has a ring-shaped structure when $\theta_\perp < V$ \citep{usami22}.
The pressures, the energy density, and the pressure anisotropy of the ring Maxwellian are as follows.
\begin{align}
P_\parallel = \frac{1}{2} N_0 m \theta_{\parallel}^2,
~~
P_\perp = \frac{1}{2} N_0 m (V^2 + \theta_{\perp}^2 ),
\\
\mathcal{E} = \frac{1}{4} N_0 m (2V^2 + \theta_{\parallel}^2 + 2\theta_{\perp}^2 ),
~~
A =
\frac{V^2 + \theta_{\perp}^2}{\theta_\parallel^2}
-1
.
\end{align}

In this study, we propose its shell-type extension,
called the shell Maxwell distribution or the shell Maxwellian.
As will be shown,
the shell Maxwellian is spherically scattered from a seed Maxwellian,
while the ring Maxwellian is axisymmetrically scattered.
Its phase space density, pressure, and energy density are given by
\begin{align}
f_\mathrm{sM}(\vec{v}) d^3v
&=
\frac{N_0}{4 rV \theta (\pi)^{3/2}}
\left\{
\exp\left( -\frac{(r-V)^2}{\theta^2} \right)
-
\exp\left( -\frac{(r+V)^2}{\theta^2} \right)
\right\}
d^3v
,
\label{eq:shellM}
\\
P &=  N_0 m \left( \frac{1}{3} V^2 + \frac{1}{2} \theta^2 \right),
~~~
\mathcal{E} = \frac{1}{4} N_0 m ( 2V^2 + 3\theta^2 )
.
\end{align}
where we use the radial speed $r=|v|=\sqrt{v_x^2+v_y^2+v_z^2}$ to emphasize its meaning in the polar coordinate system.
When $V\rightarrow 0$,
it recovers the Maxwell distribution
\begin{align}
\lim_{V\rightarrow 0} f_\mathrm{sM}(\vec{v})
=&
\frac{N_0}{2 r \theta (\pi)^{3/2}}
~
\lim_{V\rightarrow 0}
\frac{
e^{-\frac{(r-V)^2}{\theta^2}}
-
e^{-\frac{(r+V)^2}{\theta^2}}
}{2V}
=
\frac{-N_0}{2 r \theta (\pi)^{3/2}}
~
\frac{d}{dx}
e^{ -{x^2}/{\theta^2} }
\Bigg|_{x=r}
=
\frac{N_0}{(\pi \theta^2)^{3/2}}
~
e^{ -{r^2}/{\theta^2} }
\end{align}
When $\theta = 0$,
the pressure and energy density recover those of a thin shell,
\begin{align}
P = \frac{1}{3} N_0 m V^2,
~~~
\mathcal{E} = \frac{1}{2} N_0 m V^2
.
\end{align}

Here we derive Eq.~\eqref{eq:shellM}.
We first place the seed Maxwellian, whose bulk velocity is $\vec{V}$.
\begin{align}
f(\vec{v})
=
\frac{N_0}{(\pi \theta^2)^{3/2}}
\exp\left( -\frac{(\vec{v}-\vec{V})^2}{\theta^2} \right)
,
\end{align}
Then we scatter the pitch angle,
such that the distribution is fully isotropied.
Using the radial velocity $r$,
the final state $f_\mathrm{sM}(r) 4\pi r^2 dr$
should satisfy the following relation,
with help from polar coordinates of $(r, \alpha, \varphi)$ and
a pitch-angle cosine, $\mu \equiv \cos\alpha$.
\begin{align}
f_\mathrm{sM}(r) 4\pi r^2dr
=&
\frac{N_0}{(\pi \theta^2)^{3/2}}
\left\{
\iint
\exp\left( -\frac{r^2+V^2}{\theta^2} \right)
\exp\left( \frac{2rV\cos\alpha}{\theta^2} \right)
\sin\alpha
d\alpha d\varphi
\right\}
r^2 dr,
\\
=&
\frac{2\pi N_0}{(\pi \theta^2)^{3/2}}
\exp\left( -\frac{r^2+V^2}{\theta^2} \right)
\left\{
\int_{-1}^{1}
\exp\left( \frac{2rV \mu}{\theta^2} \right)
d\mu
\right\}
r^2 dr,
\end{align}
This immediately gives Eq.~\eqref{eq:shellM}.
The energy distribution resembles that of a drifting Maxwellian, discussed by \citet[][Section 2]{karlicky12}.
Maclaurin expansion of Eq.~\eqref{eq:shellM} gives further insights:
\begin{align}
f_\mathrm{sM}(r)
&=
\frac{N_0}{4 rV \theta (\pi)^{3/2}}
\exp\left( -\frac{V^2}{\theta^2} \right)
\exp\left( -\frac{r^2}{\theta^2} \right)
\left\{
\exp\left( \frac{2rV}{\theta^2} \right)
-
\exp\left( -\frac{2rV}{\theta^2} \right)
\right\}
\\
&\simeq 
\frac{N_0}{4 rV \theta (\pi)^{3/2}}
\exp\left( -\frac{V^2}{\theta^2} \right)
\left( 1 -\frac{r^2}{\theta^2} + \mathcal{O}({r^4}) \right)
\left\{
\frac{4rV}{\theta^2}
\left(
1+
\frac{2r^2V^2}{3\theta^4}
+ \mathcal{O}({r^4})
\right)
\right\}
\\
&\simeq 
\frac{N_0}{(\pi \theta^2)^{3/2}}
\exp\left( -\frac{V^2}{\theta^2} \right)
\left\{
1+
\left( \frac{2V^2-3\theta^2}{3\theta^4} \right)
r^2
+
\mathcal{O}(r^4)
\right\}
\end{align}
This indicates that
the shell Maxwellian has a peak at the center, and
it no longer looks like a shell when $V \le \sqrt{{3}/{2}}\theta \simeq 1.22\theta$.
The phase space density has a shell-shaped structure
when $V > \sqrt{{3}/{2}}\theta$.

Since they are scattered from the seed Maxwellian,
the ring and shell Maxwellians can be easily generated in PIC simulation.
One can generate a Maxwell distribution with a thermal velocity, and then
rotate the particle velocity in an axisymmetric or spherical manner. 
The procedures are presented in Algorithms 7.1 and 7.2 in Table \ref{table:shellMax}.
They are simpler than Algorithms 6.1 and 6.2
which require additional procedures. 
We have also carried out Monre-Carlo tests
for $V=5\theta_\perp$ or $V=5\theta$ with $10^6$ particles.
Figures~\ref{fig:shell}(c) and (d) compare
perpendicular and radial distributions of the distributions (black lines)
and numerical results (blue histograms).

Figure~\ref{fig:shell}(e) compares 
the ring distribution and the ring Maxwellian
for $V=5\theta$ (solid line) and for $V=\theta$ (dotted line),
and Figure~\ref{fig:shell}(f) compares 
the shell distributions in the same format.
\begin{align}
\dfrac{f_\mathrm{ring}(v_\perp) {\,2\pi v_\perp}}
{f_\mathrm{rM}(v_\perp) {\,2\pi v_\perp}}
&=
\frac{1}{C_2(\frac{V}{\theta_\perp})}
\dfrac{
\exp\left( \frac{2 v_\perp V }{\theta_\perp^2}\right)
}{
I_0\left(\frac{2 v_\perp V}{\theta_\perp^2}\right)
}
,~~~~~
\dfrac{f_\mathrm{shell}(r){\,4\pi r^2}}
{f_\mathrm{sM}(r){\,4\pi r^2}}
=
\frac{2 rV \sqrt{\pi}}{\theta^2 C_3(\frac{V}{\theta})}
\left\{
1
-
\exp\left( -\frac{4rV}{\theta^2} \right)
\right\}^{-1}
,
\label{eq:ringshell_diff}
\end{align}
As evident in Figure~\ref{fig:shell}(e),
the two ring distributions are very similar around the peaks for $V=5\theta_\perp$.
The phase space densities are maximum
at $v_\perp=5\theta_\perp$ for the ring distribution,
and at $v_\perp \approx 4.95\theta_\perp$ for the ring Maxwellian.
Also, in Figure~\ref{fig:shell}(f),
the two shell distributions are similar around the peaks for $V=5\theta$.
The maximums are located at $v=5\theta$ for the shell distribution,
and at $v \approx 4.9\theta$ for the shell Maxwellian.
These results suggest that
the ring and shell Maxwellians are good alternatives for
the ring and shell distributions with Gaussian width. 
On the other hand,
when $V \lesssim \theta$,
the ring (shell) distribution and the ring (shell) Maxwellian
are very different, as shown by the dotted lines. 
As already stated, the artificial boundary of
the ring and shell distributions becomes problematic.

\begin{table}
\centering
\caption{Algorithms for the ring and shell Maxwellians
\label{table:shellMax}}
\begin{tabular}{l}
\hline
{\bf Algorithm 7.1: ring Maxwellian}\\
\hline
generate $N_1, N_2, N_3 \sim \mathcal{N}(0,1)$ \\
generate $U \sim U(0, 1)$ \\
$v_{\parallel} ~~\leftarrow (\theta_{\parallel}/\sqrt{2})  N_1$ \\
$v_{\perp 1} \leftarrow (\theta_{\perp}/\sqrt{2})  N_2 + V \cos(2\pi U)$ \\
$v_{\perp 2} \leftarrow (\theta_{\perp}/\sqrt{2})  N_3 + V \sin(2\pi U)$ \\
{\bf return} $v_{\parallel}, v_{\perp 1}, v_{\perp 2}$\\
\hline
{\bf Algorithm 7.2: shell Maxwellian}\\
\hline
generate $N_1, N_2, N_3 \sim \mathcal{N}(0,1)$ \\
generate $U_1, U_2 \sim U(0, 1)$ \\
$v_{x} \leftarrow (\theta/\sqrt{2}) N_1 + ~ V ~ ( 2 U_1 - 1 )$ \\
$v_{y} \leftarrow (\theta/\sqrt{2}) N_2 + 2 V \sqrt{ U_1 (1-U_1) } \cos(2\pi U_2)$ \\
$v_{z} \leftarrow (\theta/\sqrt{2}) N_3 + 2 V \sqrt{ U_1 (1-U_1) } \sin(2\pi U_2)$ \\
{\bf return} $v_x, v_y, v_z$ \\
\hline
\end{tabular}
\end{table}

\section{Other isotropic distributions}
\label{sec:other}

We present numerical procedures for two isotropic distributions.
First, we discuss the super-Gaussian distribution,
also known as the self-similar distribution or the multivariate exponential power distribution.
The distribution has long been discussed in laser plasmas \citep{dum74,mora82}
but also useful in space plasmas \citep{wilson19}.
It is given by
\begin{align}
f_\mathrm{sg}(\vec{v})d^3{v}
&=
\frac{N_0p}{4 \pi \theta^3 \Gamma({3}/{p})}
\exp
\left\{
-\left(\frac{v}{\theta}\right)^p \right\}
d^3{v}
,
\label{eq:sg}
\end{align}
where $\theta$ is the characteristic velocity and $p$ is the index. 
The pressure and energy density are:
\begin{align}
P &=
\frac{1}{3} N_0 m \theta^2
\frac{\Gamma({5}/{p})}{\Gamma({3}/{p})}
,
~~~~
\mathcal{E} =
\frac{1}{2} N_0 m \theta^2
\frac{\Gamma({5}/{p})}{\Gamma({3}/{p})}
,
\end{align}

We consider the radial distribution,
\begin{align}
F_\mathrm{sg}({v}) d{v}
=
f_\mathrm{sg}({v}) 4\pi v^2 d{v}
=
\frac{N_0 p}{\theta^3 \Gamma({3}/{p})}
\exp
\left\{
-\left(\frac{v}{\theta}\right)^p \right\}
v^2
~
d{v}
,
\end{align}
Using an auxiliary variable $x\equiv (v/\theta)^p$, we find that it follows a gamma distribution,
\begin{align}
F_\mathrm{sg}(x)dx
&=
N_0
\Big\{
\mathrm{Ga}(x;{3}/{p},1)
dx
\Big\}
,
\end{align}
Thus, the velocity $v$ can be drawn from a gamma variate $X_{\mathrm{Ga}(3/p,1)}$,
\begin{align}
v = \theta \Big( X_{\mathrm{Ga}(3/p,1)} \Big)^{1/p}
\end{align}
Then one can obtain $\vec{v}$ by scattering the vector in three dimensions. 
The procedure is shown in Algorithm 8.1 (Table \ref{table:sg}).

\begin{table}
\centering
\caption{Algorithms for super-Gaussian and filled-shell distributions.
\label{table:sg}}
\begin{tabular}{l}
\hline
{\bf Algorithm 8.1: super Gaussian}\\
\hline
generate $X \sim \mathrm{Ga}(3/p, 1)$ \\
generate $U_1, U_2 \sim U(0, 1)$ \\
$v \leftarrow  \theta \, X^{1/p}$\\
$v_x \leftarrow v ~ ( 2 U_1 - 1 )$ \\
$v_y \leftarrow 2 v \sqrt{ U_1 (1-U_1) } \cos(2\pi U_2)$ \\
$v_z \leftarrow 2 v \sqrt{ U_1 (1-U_1) } \sin(2\pi U_2)$ \\
{\bf return} $v_x, v_y, v_z$ \\
\hline
{\bf Algorithm 8.2: filled-shell distribution}\\
\hline
generate $U_1, U_2, U_3 \sim U(0, 1)$ \\
$v \leftarrow  V \cdot (U_1)^{1/(3+p)} $\\
$v_x \leftarrow v ~ ( 2 U_2 - 1 )$ \\
$v_y \leftarrow 2 v \sqrt{ U_2 (1-U_2) } \cos(2\pi U_3)$ \\
$v_z \leftarrow 2 v \sqrt{ U_2 (1-U_2) } \sin(2\pi U_3)$ \\
{\bf return} $v_x, v_y, v_z$ \\
\hline
\end{tabular}
\end{table}

Finally we consider the filled-shell distribution,
which plays a role in a distant heliosphere.
The phase space density follows a power-law $\propto v^p$
inside a spherical shell ($v \le V$),
\begin{align}
f_\mathrm{fs}(\vec{v}) &= \frac{N_0(3+p)}{4\pi V^{3+p}} {v}^{p}\cdot\mathcal{H}(V-v)
\end{align}
Here, $\mathcal{H}(x)$ is the Heaviside step function and
$p$ is the spectral index ($p>-3$).
The pressure and energy density are given by
\begin{align}
P = \frac{3+p}{3(5+p)} N_0 mV^2,
~~~
\mathcal{E} = \frac{3+p}{2(5+p)} N_0 mV^2.
\end{align}
The spectral index varies from problem to problem.
The so-called Vasyliunas--Siscoe model \citep{vas76}
employs $p=-3/2$ \citep[e.g.,][]{zank10},
but we find other choices in the literature \citep[e.g., $p=1$; ][]{giacalone25}.

Since $f_\mathrm{fs}(v) 4\pi v^2 \propto v^{2+p}$,
one can generate the velocity $v = V \cdot (U_1)^{1/(3+p)}$,
using a uniform variate $U_1 \sim U(0,1)$.
A full procedure is presented in Algorithm 8.2 (Table \ref{table:sg}).

\begin{figure*}[htbp]
\centering
\includegraphics[width={\textwidth}]{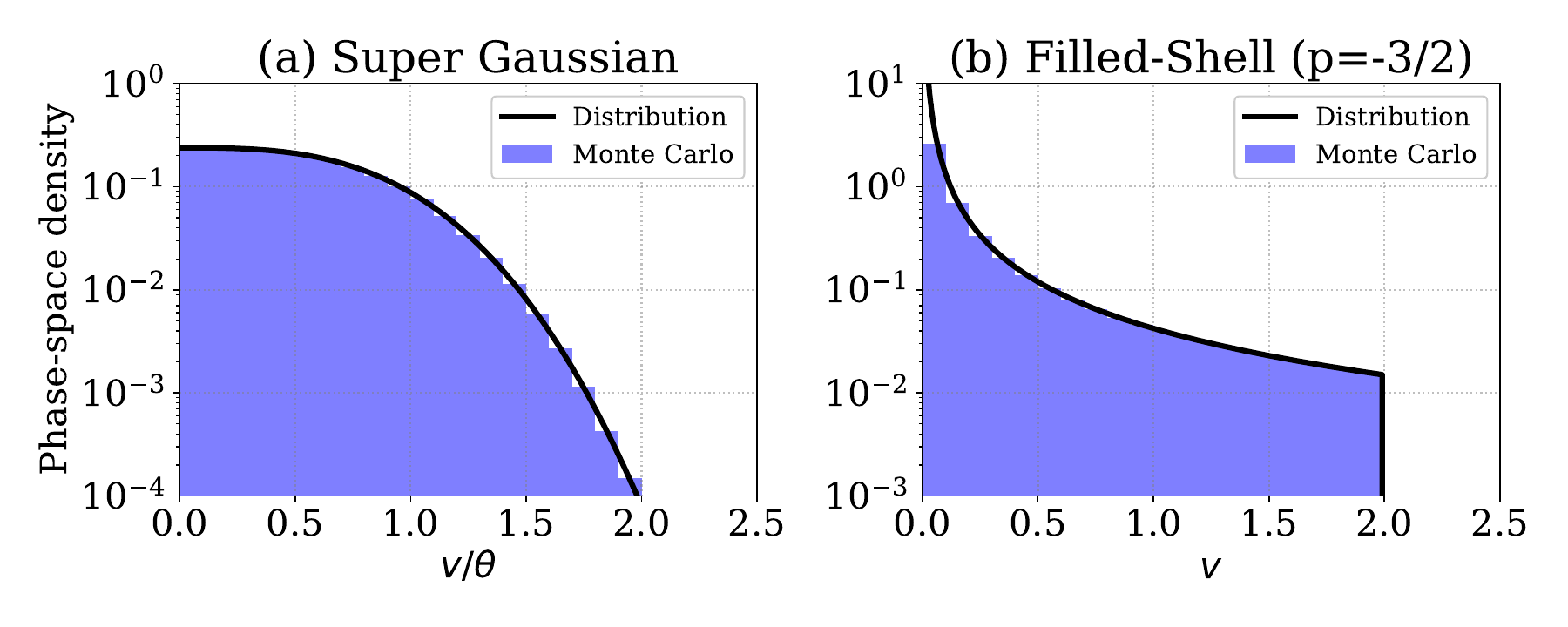}
\caption{
The phase space densities (black lines) and
Monte Carlo results of $10^6$ particles (blue histograms)
for
(a) the super Gaussian distribution ($p=3$, $\theta=1$) and
(b) the filled-shell distribution ($p=-3/2$, $V=2$)
are presented.
\label{fig:sg}}
\end{figure*}

The two procedures are numerically tested, by using $10^6$ particles.
We set $p=3$ and $\theta=1$ for the super Gaussian, and 
$p=-3/2$ and $V=2$ for the filled-shell distribution.
Their Monte Carlo results are shown in the two panels in Figure \ref{fig:sg}.
The exact solutions and numerical results are in excellent agreement.

\section{Numerical test}
\label{sec:verification}

\begin{figure}[htbp]
\centering
\includegraphics[width={0.8\columnwidth}]{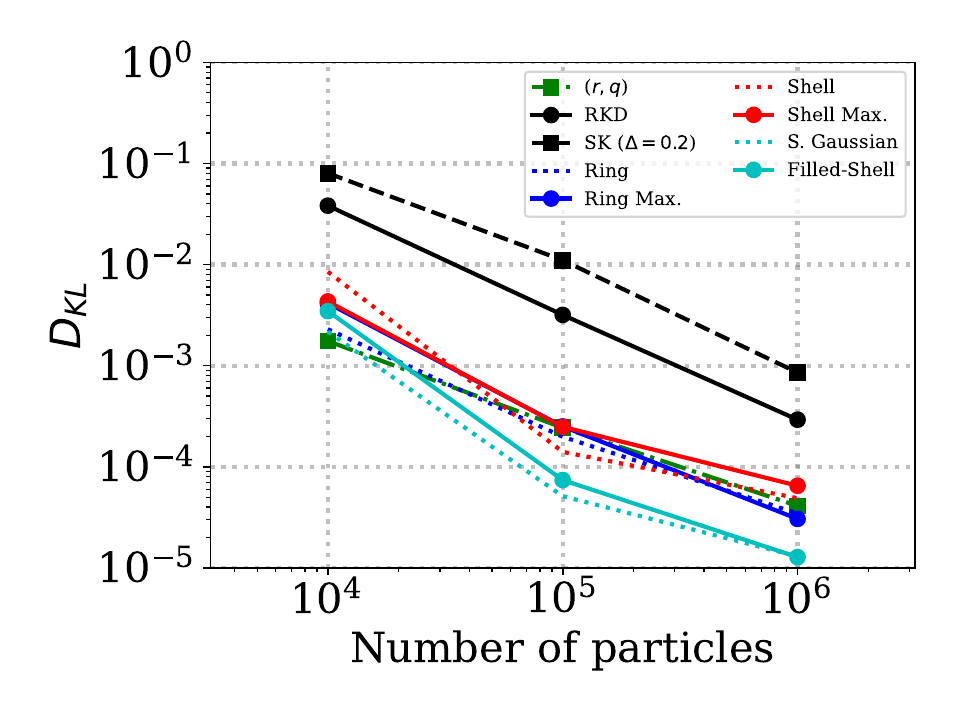}
\caption{
The Kullback--Leibler divergence of the numerical distributions from the analytical solutions (Eq.~\eqref{eq:KL}).
\label{fig:KL}}
\end{figure}

Our algorithms are tested in each sections
(Sections \ref{sec:rq}--\ref{sec:other}),
either in 1-D histograms and in the 2-D colormaps, but
we present one more test to verify them.
We have conducted a Kullback--Leibler (KL) divergence test
to our datasets, as done in our earlier study \citep{zeni23}.
For the $i$-th bin in 1-D histograms or the $i$-th cell in 2-D colormaps,
we compare exact densities $P(i)$ and
normalized particle count numbers $Q(i)$.
Similarity between $P$ and $Q$ is evaluated by the KL divergence,
also known as the cross entropy,
\begin{align}
D_{\rm{KL}}(P\|Q) \equiv \sum_i (P(i)+\epsilon) \log \frac{P(i)+\epsilon}{Q(i)+\epsilon}
\label{eq:KL}
\end{align}
where a small number $\epsilon=10^{-10}$ is added to avoid zero.
It is non-negative, and it becomes zero when and only when the two distributions are identical.

Fig.~\ref{fig:KL} shows the KL divergence
for the nine distributions,
as a function of the total particle number.
The parameters are detailed in the relevant sections.
For the subtracted Kappa distribution,
$\Delta=0.2$ is set (Fig.~\ref{fig:SK}(b)).
For the ring and shell distributions (Figs.~\ref{fig:shell}(a,b)),
the piecewise rejection method is used.
Note that it is not meaningful to compare different distributions quantitatively,
because the data come from sources of difference sizes:
PSDs, omni-directional velocity distributions ($4\pi v^2 f(v)$), or a 2-D colormap.
Importantly, $D_{\rm{KL}}$ approaches zero in all the cases,
as we increase the particle number.
This suggests that
the Monte Carlo results approximate the exact distributions very well.

\section{Discussion and Summary}
\label{sec:discussion}

In our earlier papers \citep{zeni22,zeni23}, 
we have proposed numerical procedures using three elemental variates:
the uniform variate, the normal variate, and the gamma variate.
In this paper, we have successfully extended
this approach to nine different VDFs:
the $(r,q)$ distribution, 
the regularized Kappa distribution,
the subtracted Kappa distribution,
the ring distribution,
the shell distribution,
the ring Maxwellian,
the shell Maxwellian,
the super-Gaussian, and 
the filled-shell distributions.
With help from the basic statistical distributions (Section \ref{sec:prep}),
mathematics behind the recipes are fully explained.
The numerical recipes for all the VDFs are provided in Tables,
making it easy for users to implement them in the code. 
The algorithms are tested in each sections,
as well as in Section \ref{sec:verification}.

In Section \ref{sec:rq},
we have discussed the $(r,q)$ distribution that generalizes
the Kappa and flattop distributions.
One can use either of
the beta-prime method and the piecewise rejection method,
if one is aware of limitations of each. 
Sections \ref{sec:RKD} and \ref{sec:SK} have discussed
the two advanced Kappa distributions. 
For the regularized Kappa distribution,
the two rejection methods and their preferred parameter spaces are described. 
The simple procedure was proposed for
a very recent model of the subtracted Kappa distribution.
The loss-cone filling parameter $\Delta$ is considered in Eq.~\eqref{eq:subM_x2}.
A previous procedure for the subtracted Maxwellian \citep[][Section 2]{zeni23}
can be similarly modified to incorporate the filling parameter. 
In addition, after submitting this paper,
\citet{min25} independently developed the same procedure for their PIC simulation
(see Appendix in their paper).
This work provides a more detailed explanation to this cutting-edge method.

In Section \ref{sec:ringMax},
we have introduced the ring Maxwellian and the shell Maxwellian.
We would propose them as alternatives for
the ring and shell distributions in Section \ref{sec:ring}. 
When the ring/shell velocity substantially exceeds the thermal width, $V \gg \theta$,
these distributions are similar to the conventional ring and shell distributions.
If we evaluate their differences using Eq.~\eqref{eq:ringshell_diff} at 
$V\pm \sigma$, where $\sigma  = \theta/\sqrt{2}$ is the standard deviation,
the two ring distributions are in agreement within 10\% for $V \gtrsim 3.9\theta_\perp$,
and the two shell distributions are in agreement within 15\% for $V \gtrsim 5.25\theta$.
Physically, the new Maxwellians correspond to
the radial or spherical scattering of a seed population,
whereas the ring and shell distributions lack a justification for their Gaussian width. 
When $V \lesssim \theta$, the conventional distributions become somewhat artificial near $v=0$, and probably it does not make sense to employ them.
In contrast, the (ring and shell) Maxwellians have no artificial problem at $v=0$, but they are no longer ring- or shell-shaped
when $V \le \theta_{\perp}$ (ring) or $\le \sqrt{3/2} \theta$. 
Practically, the new Maxwellians are easier to use.
As shown in Sections \ref{sec:ring} and \ref{sec:ringMax},
it is much easier to estimate
pressures, pressure anisotropies, and energy density for the new Maxwellians
than those for the conventional distributions.
Also, Monte Carlo procedures for the new Maxwellians are
much simpler than those for the conventional distributions,
as shown in Tables \ref{table:ringshell} and \ref{table:shellMax}. 
These facts suggest that
the ring or shell Maxwellians are favorable for
modeling of heliospheric kinetic processes in ring or shell-shaped VDFs.

Technically,
the proposed algorithms raise minor concerns
when deployed on graphics processing units (GPUs).
In particular, algorithms that rely on the rejection method
can be inefficient on GPUs.
This is because the number of iterations required by the rejection loop may vary from thread to thread, which is not well suited to the GPU's execution model.
Furthermore, at the time of writing,
neither NVIDIA’s {\tt cuRAND} library nor AMD’s {\tt hipRAND} library provides gamma generators,
although both libraries support uniform and normal variates.
Consequently, we need to generate gamma variates on the CPU or
to implement custom gamma generators on the GPU.
Since GPU computation has only recently become popular in heliophysics, further investigation is needed to see whether these issues are critical and to find other potential problems.

Together with our earlier works \citep{zeni22,zeni23},
we can now generate a wide variety of 
single-component VDFs in heliophysics for particle simulations. 
In more complex senarios, VDFs may have multiple components,
some of which are not approximated by well-known VDFs. 
For example, solar-wind electrons often
consists of core, halo, and strahl populations. 
In such cases, it may be necessary
not only to develop separate procedures for each component,
but also to use a fully numerical inversion method \citep{an22}. 
This should be evaluated on a case-by-case basis. 

In summary, we have provided numerical procedures
for generating nine different VDFs in heliophysics research.
We hope that this work will make
kinetic modeling of heliospheric plasma processes easier than before.

\section*{Acknowledgements}
SU was supported by Grant-in-Aid for Scientific Research (C) 22K03581 from the Japan Society for the Promotion of Science (JSPS).
SM was supported by Grant-in-Aid for Scientific Research (B) 23K22558 from JSPS.

\section*{Author Declarations}
\subsection*{Conflict of Interest}
The authors have no conflicts to disclose.

\subsection*{Author Contributions}
{\bf SZ}: Conceptualization (lead), Formal analysis (lead), Investigation (equal), Methodology (lead), Visualization (lead), Writing – original draft (lead), Writing – review \& editing (equal), 
{\bf SU}: Conceptualization (supporting), Investigation (equal), Writing – review \& editing (equal),
{\bf SM}: Conceptualization (supporting), Investigation (equal), Writing – review \& editing (equal)

\subsection*{Data Availability Statement}
The Jupyter notebook for this article is archived in Zenodo \citep{zeni26}.

\appendix

\section{Piecewise rejection method for log-concave distributions}
\label{sec:appendix}

We first rewrite the target distributions.
\begin{align}
g(v) =
\left\{
\begin{array}{ll}
g_\mathrm{ring}(v_\perp)
& \mathrm{(Eq.~\eqref{eq:ring_g}; ring~distribution)}
\\\\
g_\mathrm{shell}(v)
& \mathrm{(Eq.~\eqref{eq:shell_g}; shell~distribution)}
\end{array}
\right.
\label{eq:ringshell_g}
\end{align}
The distribution has a maximum at the most probable speed: 
\begin{align}
v_m =
\left\{
\begin{array}{ll}
\frac{1}{2}\left( V+\sqrt{V^2+2\theta_\perp^2} \right)
~~~~
& \mathrm{(ring)}
\\
\frac{1}{2}\left( V+\sqrt{V^2+4\theta^2} \right)
& \mathrm{(shell)}
\end{array}
\right.
\label{eq:ringshell_mp}
\end{align}
With this in mind, we define left and right contact points,
\begin{align}
v_L \equiv v_m - \theta,~~
v_R \equiv v_m + \theta
\label{eq:LR}
\end{align}
To ensure $v_L > 0$, $V > \theta/2$ ~($V > 0$) is required
in the case of the ring (shell) distribution.
This is reasonable limitation, because we usually consider $V \gtrsim \theta$. 

Since $g(x)$ is log-concave,
i.e., $\big( \log g_\mathrm{ring}({v}_\perp) \big)'' \propto - ({1}/{v_\perp^2} + {2}/{\theta_\perp^2}) < 0$
and
$( \log g_\mathrm{shell}(v) )'' \propto -(2/v^2 + 2/\theta^2) < 0$,
we can define exponential envelope functions that
touch the distribution at $v = v_L, v_m$, and $v_R$.
\begin{align}
G(p)
&=
\left\{
\begin{array}{ll}
g(v_L) \exp\left(\dfrac{v-v_L}{\lambda(v_L)}\right)
~~
& (0 \le v < x_L)
\\ 
g(v_m)
~~
& (x_L \le v < x_R)
\\ 
g(v_R) \exp\left(-\dfrac{v-v_R}{\lambda(v_R)}\right)
& (x_R \le v < \infty)
\end{array}
\right.
\label{eq:envelope1}
\end{align}
Here, $\lambda(v) = \left| {g(v)}/{g'(v)} \right|$
is the length scale and
\begin{align}
x_L = v_L - \lambda(v_L) \log \frac{g(v_L)}{g(v_m)},~~
x_R = v_R + \lambda(v_R) \log \frac{g(v_R)}{g(v_m)}
\label{eq:swisdak_xR}
\end{align}
are the two switching points.
As an example,
the envelope functions are drawn by the dashed orange lines
in Figures \ref{fig:shell}(a,b). 
The ratio of the areas below the envelopes is
$\lambda(v_L) : (x_R - x_L ) : \lambda(v_R)$.
Then a rejection algorithm can be constructed,
as shown in Table~\ref{table:swisdak}.
The acceptance efficiency is
\begin{align}
{\rm eff} =
\frac{1}{ g(v_m) \big\{ (x_R-x_L) + \lambda( v_R ) + \lambda( v_L ) \big\} }
.
\end{align}
This gives
$\gtrsim 86\%$ for $V \gtrsim \theta$ and 
$88\%$--$89\%$ for $V \gtrsim 2\theta$
for both ring and shell distributions. 
In practice, the normalization constant in $g(x)$ can be dropped
in the procedure (Table~\ref{table:swisdak}).

\begin{table}[tpb]
\centering
\caption{Piecewise rejection method
\label{table:swisdak}}
\begin{tabular}{l}
\hline
{\bf Piecewise rejection method for Algorithms 6.1 and 6.2}\\
\hline
$V > \theta_\perp/2$ is required for the ring distribution \\
$g(v)$ is defined by Eq.~\eqref{eq:ringshell_g} (Eq.~\eqref{eq:ring_g} or \eqref{eq:shell_g})\\
$v_m$ is defined by Eq.~\eqref{eq:ringshell_mp} \\
\hline
$g_{max} \leftarrow g(v_m)$ \\
$v_L \leftarrow v_m - \theta$,~
$\lambda_L \leftarrow \dfrac{g(v_L)}{g'(v_L)} $,~
$x_L \leftarrow v_L - \lambda_L \log \dfrac{g(v_L)}{g_{max}}$\\
$v_R \leftarrow v_m + \theta$,~
$\lambda_R \leftarrow \Big| \dfrac{g(v_R)}{g'(v_R)} \Big|$,~
$x_R \leftarrow v_R + \lambda_R \log \dfrac{g(v_R)}{g_{max}}$
\\
$S \leftarrow \lambda_L + (x_R-x_L) + \lambda_R$,~\\
$p_L \leftarrow \dfrac{\lambda_L}{S}$,~
$p_R \leftarrow \dfrac{\lambda_R}{S}$,~
$p_C \leftarrow 1 - p_L - p_R$\\
{\bf repeat}\\
~~~~generate $U_1, U_2 \sim U(0, 1)$ \\
~~~~{\bf if}~$U_1 \le p_C$ {\bf then} \\
~~~~~~~~$v \leftarrow x_L + (x_R-x_L) (U_1/p_C)$ \\
~~~~~~~~{\bf if}~$U_2 < g(v)/g_{max}$ {\bf break} \\
~~~~{\bf else if}~$U_1 \le p_C + p_L$ {\bf then} \\
~~~~~~~~$u \leftarrow ({U_1-p_C})/{p_L}$ \\
~~~~~~~~$v \leftarrow x_L + \lambda_L\log u$ \\
~~~~~~~~{\bf if}~$0<v$ {\bf and}~$u U_2 < g(v)/g_{max}$ {\bf break} \\
~~~~{\bf else}  \\
~~~~~~~~$u \leftarrow (U_1-p_C-p_L)/{p_R} $ \\
~~~~~~~~$v \leftarrow x_R - \lambda_R \log u$ \\
~~~~~~~~{\bf if}~$u U_2 < g(v)/g_{max}$ {\bf break} \\
~~~~{\bf endif}\\
{\bf end repeat}
\\
{\bf return} $v$
\\
\hline
\end{tabular}
\end{table}



\begin{thebibliography}{}
\bibitem[{\textit{Abdul \& Mace}}(2014)]{abdul14}
Abdul R. F., \& R. L. Mace (2014), A method to generate kappa distributed random deviates for particle-in-cell simulations, \cpc {\itshape 185}, 2383--2386, doi:10.1016/j.cpc.2014.05.006.
\bibitem[{\textit{Abdul \& Mace}}(2015)]{abdul15}
Abdul R. F., \& R. L. Mace (2015), One-dimensional particle-in-cell simulations of electrostatic Bernstein waves in plasmas with kappa velocity distributions, \pop {\itshape 22}, 102107, doi:10.1063/1.4933005.
\bibitem[{\textit{Ahrens \& Dieter}}(1974)]{AD74}
Ahrens, J. H., \& U. Dieter (1974), Computer Methods for Sampling from Gamma, Beta, Poisson and Binomial Distributions, {\itshape Computing}, {\itshape 12}, 223--246, doi:10.1007/BF02293108.
\bibitem[\textit{An et al.}(2022)]{an22}
An, X., A. Artemyev, V. Angelopoulos, S. Lu, P. Pritchett, V. Decyk (2022),
Fast inverse transform sampling of non-Gaussian distribution functions in space plasmas. \jgr {\itshape 127}, e2021JA030031, doi:10.1029/2021JA030031.
\bibitem[{\textit{Ashour-Abdalla \& Kennel}}(1978)]{AK78}
Ashour-Abdalla, M., \& C. F. Kennel (1978), Nonconvective and convective electron cyclotron harmonic instabilities, \jgr {\itshape 83}, 1531--1543, doi:10.1029/JA083iA04p01531.
\bibitem[{\textit{Best}}(1983)]{best83}
Best, D. J. (1983), A Note on Gamma Variate Generators with Shape Parameter less than Unity, {\itshape Computing}, {\itshape 30}, 185--188, doi:10.1007/BF02280789.
\bibitem[{\textit{Birdsall \& Langdon}}(1985)]{birdsall}
Birdsall, C.~K., \& A. B. Langdon (1985), {\itshape Plasma Physics via Computer Simulation}, McGraw-Hill, New York.
\bibitem[{\textit{Box--Muller}}(1958)]{bm58}
Box G. E. P., \& M. E. Muller (1958), A note on the generation of random normal deviates, {\itshape Ann. Math. Stat.}, {\itshape 29}, 610--611, doi:10.1214/aoms/1177706645.
\bibitem[{\textit{Devroye}}(1986)]{devroye86}
Devroye, L. (1986), {\itshape Non-Uniform Random Variate Generation}, Springer-Verlag, available at http://luc.devroye.org/rnbookindex.html.
\bibitem[{\textit{Dory et al.}}(1965)]{DGH}
Dory, R. A., G. E. Guest, \& E. G. Harris (1965), Unstable Electrostatic Plasma Waves Propagating Perpendicular to a Magnetic Field, \prl {\itshape 14}, 131--133, doi:10.1103/PhysRevLett.14.131.
\bibitem[{\textit{Dum et al.}}(1974)]{dum74}
Dum, C. T., R. Chodura, \& D. Biskamp (1974), Turbulent Heating and Quenching of the Ion Sound Instability, \prl {\bf 32}, 1231, doi:10.1103/PhysRevLett.32.1231.
\bibitem[{\textit{Feldman et al.}}(1982)]{feldman82}
Feldman, W. C., S. J. Bame, S. P. Gary, J. T. Gosling, D. McComas, \& M. F. Thomsen (1982), Electron Heating Within the Earth's Bow Shock, \prl, 49, 199--201, doi:10.1103/PhysRevLett.49.199.
\bibitem[{\textit{Freund \& Wu}}(1988)]{freund88}
Freund, H. P., \& C. S. Wu (1988), Stability of a spherical shell distribution of pickup ions, \jgr {\itshape 93}, 14277--14283, doi:10.1029/JA093iA12p14277.
\bibitem[{\textit{Giacalone et al.}}(2025)]{giacalone25}
Giacalone, J., M. Kornbleuth, M. Opher, M. Gkioulidou, J. K{\o}ta, E. Puzzoni, J. D. Richardson, \& G. P. Zank (2025),
Hybrid Simulations of Interstellar Pickup Ions at the Solar Wind Termination Shock Revisited,
\apj {\itshape 980}, 29, doi:10.3847/1538-4357/ada89c.
\bibitem[{\textit{Hockney \& Eastwood}}(1981)]{hockney}
Hockney, R.~W., \& J.~W. Eastwood (1981), {\itshape Computer simulation using particles}, McGraw-Hill, New York.
\bibitem[{\textit{Karlick\'{y} et al.}}(2012)]{karlicky12}
Karlick\'{y}, M., E. Dzif\v{c}\'{a}kov\'{a}, \& J. Dud\'{i}k (2012), On the physical meaning of $n$-distributions in solar flares, {\itshape Astron. Astrophys., 537}, A36, doi:10.1051/0004-6361/201117860.
\bibitem[{\textit{Kennel}}(1966)]{kennel66}
Kennel, C. F. (1966), Low-Frequency Whistler Mode, {\itshape Phys. Fluids}, {\itshape 9}, 2190--2202, doi:10.1063/1.1761588.
\bibitem[{\textit{Kroese et al.}}(2011)]{kroese11}
Kroese, D. P., T. Taimre, \& Z. I. Botev (2011), {\itshape Handbook of Monte Carlo methods}, John Wiley \& Sons.
\bibitem[{\textit{Kumar et al.}}(1997)]{kumar97}
Kumar, S., S. K. Dixit, \& A. K. Gwal (1997), Electron cyclotron waves in the presence of parallel electric fields in the Earth's auroral plasma, {\itshape Ann. Geophys., 15}, 24--28, doi:10.1007/s00585-997-0024-3.
\bibitem[{\textit{Lazar \& Fichtner}}(2021)]{lazar21}
Lazar, M., \& Fichtner, H. (eds.) (2021), {\itshape Kappa Distributions; From Observational Evidences via Controversial Predictions to a Consistent Theory of Nonequilibrium Plasmas,} Astrophysics and Space Science Library, vol. 464, Berlin: Springer, ISBN: 978-3-030-82623-9.
\bibitem[{\textit{Liu et al.}}(2011)]{liu11}
Liu, K., S. P. Gary, \& D. Winske (2011), Excitation of magnetosonic waves in the terrestrial magnetosphere: Particle-in-cell simulations, \jgr {\itshape 116}, A07212, doi:10.1029/2010JA016372.
\bibitem[{\textit{Livadiotis}}(2017)]{kappa}
Livadiotis, G. (ed.) (2017), {\itshape Kappa Distributions: Theory and Applications in Plasmas,} Elsevier, Amsterdam.
\bibitem[{\textit{Marsaglia \& Tsang}}(2000)]{mt00}
Marsaglia G., \& W. W. Tsang (2000), A simple method for generating gamma variables, {\itshape ACM Trans. Math. Software}, {\itshape 26}, 363--372, doi:10.1145/358407.358414.
\bibitem[{\textit{Matsukiyo \& Scholer}}(2014)]{matsukiyo14}
Matsukiyo, S., \& M. Scholer (2014), Simulations of pickup ion mediated quasiperpendicular shocks: Implications for the heliospheric termination shock, \jgr {\itshape 119}, 2388--2399, doi:10.1002/2013JA019654.
\bibitem[{\textit{Min \& Liu}}(2015)]{min15}
Min, K., \& K. Liu (2015), Fast magnetosonic waves driven by shell velocity distributions, \jgr {\itshape 120}, 2739--2753, doi:10.1002/2015JA021041.
\bibitem[{\textit{Min et al.}}(2025)]{min25}
Min, K., Miyoshi, Y., \& K. Liu (2025), Linear analysis and PIC simulations of electron cyclotron harmonic instability driven by a subtracted-kappa velocity distribution, \pop {\itshape 32}, 122901, doi:10.1063/5.0305058.
\bibitem[\textit{M\"{o}bius et al.}(1985)]{moebius85}
M\"{o}bius, E., D. Hovestadt, B. Klecker, M. Scholer, G. Gloeckler, \& F. M.
Ipavich (1985), Direct observation of He${}^{+}$ pick-up ions of interstellar origin in the solar wind, {\itshape Nature 318}, 426--429, doi:10.1038/318426a0.
\bibitem[{\textit{Mora \& Yahi}}(1982)]{mora82}
Mora, P., \& H. Yahi (1982), Thermal heat-flux reduction in laser-produced plasmas, {\itshape Phys. Rev. A,} {\itshape 26}, 2259, doi:10.1103/PhysRevA.26.2259.
\bibitem[\textit{Nakanotani et al.}(2023)]{nakanotani23}
Nakanotani M., G. P. Zank, \& L.-L. Zhao (2023), Pickup Ion-Mediated Magnetic Reconnection in the Outer Heliosphere, \apj {\itshape 949}, L2, doi:10.3847/2041-8213/acd33f.
\bibitem[{\textit{Oka et al.}}(2022)]{oka22}
Oka, M., T. D. Phan, M. {\O}ieroset, D. L. Turner, J. F. Drake, X. Li, S. A. Fuselier, D. J. Gershman, B. L. Giles, R. E. Ergun, R. B. Torbert, H. Y. Wei, R. J. Strangeway, C. T. Russell, \& J. L. Burch (2022), Electron energization and thermal to non-thermal energy partition during earth’s magnetotail reconnection, \pop {\itshape 29}, 052904, doi:10.1063/5.0085647.
\bibitem[{\textit{Olbert}}(1968)]{olbert68}
Olbert, S. (1968), Summary of Experimental Results from M.I.T. Detector on IMP-1. In: Carovillano, R. L., McClay, J. F., Radoski, H. R. (Eds.) {\itshape Physics of the Magnetosphere} (pp. 641-659). Astrophysics and Space Science Library, vol. 10, Springer, Dordrecht, doi:10.1007/978-94-010-3467-8\_23.
\bibitem[{\textit{Olver et al.}}(2025)]{dlmf}
Olver, F. W. J., A.~B. {Olde Daalhuis}, D.~W. Lozier, B.~I. Schneider, R.~F. Boisvert, C.~W. Clark, B.~R. Miller, B.~V. Saunders, H.~S. Cohl, and M.~A. McClain. 2025, {\itshape NIST Digital Library of Mathematical Functions}, https://dlmf.nist.gov/ (DLMF), Release 1.2.5 of 2025-12-15.
\bibitem[{\textit{Pyakurel et al.}}(2025)]{pyakurel25}
Pyakurel, P. S., M. Swisdak, S. Eriksson, B. L. Shrestha, Y.-H. Liu, J. M. TenBarge, M. A. Shay, and T. D. Phan (2025), Kinetic Simulation of Magnetic Reconnection with Pickup Ions: Shell Stability and Reconnection Rate Analysis, \apj {\itshape 988}, 119, doi:10.3847/1538-4357/ade3db.
\bibitem[{\textit{Qureshi et al.}}(2014)]{qureshi14}
Qureshi, M. N. S., W. Nasir, W. Masood, P. H. Yoon, H. A. Shah, \& S. J. Schwartz (2014), Terrestrial ion roars and non-Maxwellian distribution, \jgr {\itshape 119}, 10,059--10,067, doi:10.1002/2014JA020476.
\bibitem[{\textit{Qureshi et al.}}(2019)]{qureshi19}
Qureshi, M. N. S., W. Nasir, R. Bruno, \& W. Masood (2019), Whistler instability based on observed flat-top two-component electron distributions in the Earth's magnetosphere, \mnras {\itshape 488}, 954--964, doi:10.1093/mnras/stz1702.
\bibitem[{\textit{Qureshi et al.}}(2004)]{qureshi04}
Qureshi, M. N. S., H. A. Shah, G. Murtaza, S. J. Schwartz, \& F. Mahmood (2004), Parallel propagating electromagnetic modes with the generalized (r,q) distribution function, \pop {\itshape 11}, 3819, doi:10.1063/1.1688329. 
\bibitem[{\textit{Richard et al.}}(2025)]{richard25}
Richard, L., Y. Khotyaintsev, C. Norgren, K. Steinvall, D. Graham, J. Egedal, A. Vaivads, \& R. Nakamura (2025). Electron Heating by Parallel Electric Fields in Magnetotail Reconnection, \prl {\itshape 134}, 215201, doi:10.1103/PhysRevLett.134.215201.
\bibitem[{\textit{Scherer et al.}}(2017)]{scherer17}
Scherer, K., H. Fichtner, \& M. Lazar (2017), Regularized $\kappa$-distributions with non-diverging moments, {\itshape Europhys. Lett.} {\bf 120}, 50002, doi:10.1209/0295-5075/120/50002.
\bibitem[{\textit{Scherer et al.}}(2020)]{scherer20}
Scherer, K., E. Husidic, M. Lazar, \& H. Fichtner (2020), The $\kappa$-cookbook: a novel generalizing approach to unify $\kappa$-like
distributions for plasma particle modelling, \mnras {\bf 497}, 1738--1756, doi:10.1093/mnras/staa1969.
\bibitem[{\textit{Summers \& Stone}}(2025)]{summers25}
Summers, D., \& S. Stone (2025), The subtracted-kappa distribution and its properties, \pop {\itshape 32}, 012112, doi:10.1063/5.0239741.
\bibitem[{\textit{Summers \& Thorne}}(1991)]{summers91}
Summers, D., \& R. M. Thorne (1991), The modified plasma dispersion function, {\itshape Phys. Fluids B}, {\itshape 3}, 1835--1847, doi:10.1063/1.859653.
\bibitem[{\textit{Thomsen et al.}}(1983)]{thomsen83}
Thomsen, M. F., H. C. Barr, S. P. Gary, W. C. Feldman, \& T. E. Cole (1983), Stability of Electron Distributions Within the Earth's Bow Shock, \jgr {\itshape 88}, 3035, doi:10.1029/JA088iA04p03035.
\bibitem[{\textit{Tsallis}}(2023)]{tsallis23}
Tsallis, C. (2023), {\itshape Introduction to Nonextensive Statistical Mechanics: Approaching a Complex World, 2nd edition}, Springer, ISBN:978-3-030-79569-6.
\bibitem[{\textit{Umeda et al.}}(2007)]{umeda07}
{Umeda}, T., M. {Ashour-Abdalla}, D. {Schriver}, R. L. {Richard}, \& F. V. {Coroniti} (2007), Particle-in-cell simulation of Maxwellian ring velocity distribution, \jgr {\itshape 112}, A04212, doi:10.1029/2006JA012124.
\bibitem[{\textit{Usami \& Horiuchi}}(2022)]{usami22}
Usami, S., \& R. {Horiuchi} (2022), Pseudo-Maxwellian Velocity Distribution Formed by the Pickup-like Process in Magnetic Reconnection, {\itshape {Frontiers in Astronomy and Space Sciences}}, {\itshape 9},
{846395}, doi:{10.3389/fspas.2022.846395}.
\bibitem[{\textit{Vasyliunas}}(1968)]{vas68}
Vasyliunas, V. M. (1968), A survey of low-energy electrons in the evening sector of the magnetosphere with OGO 1 and OGO 3, \jgr {\itshape 73}, 2839--2884, doi:10.1029/JA073i009p02839.
\bibitem[{\textit{Vasyliunas \& Siscoe}}(1976)]{vas76}
Vasyliunas, V. M., \& G. L. Siscoe (1976), On the Flux and the Energy Spectrum of Interstellar Ions in the Solar System, \jgr {\itshape 81}, 1247--1252, doi:10.1029/JA081i007p01247.
\bibitem[{\textit{Wilson et al.}}(2019)]{wilson19}
Wilson III, L. B.,
L.-J. Chen, S. Wang, S. J. Schwartz, D. L. Turner, M. L. Stevens, J. C. Kasper, A. Osmane, D. Caprioli, S. D. Bale, M. P. Pulupa, C. S. Salem, \& K. A. Goodrich (2019),
Electron Energy Partition across Interplanetary Shocks. I. Methodology and Data
Product, \apjs {\itshape 243}, 8 doi:10.3847/1538-4365/ab22bd.
\bibitem[{\textit{Wu et al.}}(1989)]{wu89}
Wu, C. S., P. H. Yoon, \& H. P. Freund (1989), A theory of electron cyclotron waves generated along auroral field lines observed by ground facilities, \grl {\itshape 16}, 1461 doi:10.1029/GL016i012p01461.
\bibitem[{\textit{Xiao et al.}}(1998)]{xiao98}
Xiao, F., R. M. Thorne, \& D. Summers (1998), Instability of electromagnetic R-mode waves in a relativistic plasma, \pop {\itshape 7}, 2489--2497, doi:10.1063/1.872932.
\bibitem[{\textit{Yoon et al.}}(1989)]{yoon89}
Yoon, P. H., M. E. Mandt, \& C. S. Wu (1989), Evolution of an unstable shell distribution of pickup cometary ions, \grl {\itshape 16}, 1473--1476, doi:10.1029/GL016i012p01473.
\bibitem[{\textit{Yotsuji}}(2010)]{yotsuji10}
Yotsuji, T. (2010), {\itshape Random number generation of probability distributions for computer simulations,} Pleiades Publishing, Nagano [in Japanese].
\bibitem[{\textit{Zaheer et al.}}(2004)]{zaheer04}
Zaheer, S., G. Murtaza, \& H. A. Shah (2004), Some electrostatic modes based on non-Maxwellian distribution functions, \pop {\itshape 11}, 2246, doi:10.1063/1.1688330.
\bibitem[{\textit{Zank et al.}}(2010)]{zank10}
Zank, G. P., J. Heerikhuisen, N. V. Pogorelov, R. Burrows, \& D. McComas (2010), Microstructure Of The Heliospheric Termination Shock: Implications For Energetic
Neutral Atom Observations, \apj {\itshape 708}, 1092, doi:10.1088/0004-637X/708/2/1092.
\bibitem[{\textit{Zenitani \& Nakano}}(2022)]{zeni22}
Zenitani, S., \& S. Nakano (2022), Loading a relativistic Kappa distribution in particle simulations, \pop {\itshape 29}, 113904, doi:10.1063/5.0117628.
\bibitem[{\textit{Zenitani \& Nakano}}(2023)]{zeni23}
Zenitani, S., \& S. Nakano (2023), Loading loss-cone distributions in particle simulations, \jgr {\itshape 128}, e2023JA031983, doi:10.1029/2023JA031983.
\bibitem[{\textit{Zenitani}}(2024a)]{zeni24a}
{Zenitani}, S. (2024a), A note on the flattop velocity distribution in space plasmas, {\itshape Research Notes of the AAS}, {\itshape 8}, 30, doi:10.3847/2515-5172/ad225c.
\bibitem[{\textit{Zenitani}}(2024b)]{zeni24b}
{Zenitani}, S. (2024b), A gamma variate generator with shape parameter less than unity, {\itshape Economics Bulletin}, {\itshape 44}, 1113, arXiv:2411.01415.
\bibitem[{\textit{Zenitani}}(2025)]{zeni25}
{Zenitani}, S. (2025), A simple procedure for generating a Kappa distribution in PIC simulation, {\itshape Research Notes of the AAS}, {\itshape 9}, 299, doi:10.3847/2515-5172/ae1c41.
\bibitem[{\textit{Zenitani et al.}}(2026)]{zeni26}
{Zenitani}, S., S. Usami, \& S. Matsukiyo (2026), Loading non-Maxwellian Velocity Distributions in Particle Simulations [Software], {\itshape Zenodo}, https://doi.org/10.5281/zenodo.17148070.
\bibitem[{\textit{Zirnstein et al.}}(2022)]{zirnstein22}
Zirnstein, E.~J., {M{\"o}bius}, E., {Zhang}, M., {Bower}, J., {Elliott}, H.~A., {McComas}, D.~J., {Pogorelov}, N.~V., \& {Swaczyna}, P. (2022), In Situ Observations of Interstellar Pickup Ions from 1 au to the Outer Heliosphere, \ssr 218, 28, doi:10.1007/s11214-022-00895-2.

\end{thebibliography}
\end{document}